\documentclass{article}

\usepackage[final]{neurips_2022}
\usepackage{url}            
\usepackage{amsmath}
\usepackage{mdframed}
\usepackage{tcolorbox}
\usepackage{xcolor}
\usepackage{placeins}
\usepackage{booktabs}
\usepackage{amssymb}
\usepackage{float}

\newcommand{\revised}[1]{{\color{black}#1}}

\title{Toward a realistic model of speech processing in the brain with self-supervised learning}

\author{Juliette Millet$^{* \ 1,2,3}$ \quad Charlotte Caucheteux$^{* \ 1,4}$ \quad Pierre Orhan$^{2}$ \quad Yves Boubenec$^{2}$ \\ \textbf{Alexandre Gramfort}$^{4}$ \quad \textbf{Ewan Dunbar}$^{2,5}$ \quad \textbf{Christophe Pallier}$^{6}$ \quad \textbf{Jean-Rémi King}$^{1,2}$ \\
\\
$^{*}$These authors contributed equally \\ $^1$Meta AI, Paris, France \quad $^2$Ecole Normale Supérieure, PSL University, Paris, France \\ $^3$LPI, Université de Paris cité, Paris, France \\ $^4$Université Paris-Saclay, Inria, CEA, Palaiseau, France \\ $^5$University of Toronto, Toronto, Canada \\ $^6$Cognitive Neuroimaging Unit, INSERM, Gif-sur-Yvette, France\\
\\
\texttt{\{jumi,ccaucheteux,jeanremi\}@meta.com}
}

\begin{document}

\maketitle

\begin{abstract} 
Several deep neural networks have recently been shown to generate activations similar to those of the brain in response to the same input. These algorithms, however, remain largely implausible: they require (1) extraordinarily large amounts of data, (2) unobtainable supervised labels, (3) textual rather than raw sensory input, and / or (4) implausibly large memory (e.g. thousands of contextual words). These elements highlight the need to identify algorithms that, under these limitations, would suffice to account for both behavioral and brain responses. Focusing on speech processing, we here hypothesize that self-supervised algorithms trained on the raw waveform constitute a promising candidate. Specifically, we compare a recent self-supervised model, wav2vec 2.0, to the brain activity of 412 English, French, and Mandarin individuals recorded with functional Magnetic Resonance Imaging (fMRI), while they listened to approximately one hour of audio books. First, we show that this algorithm learns brain-like representations with as little as 600 hours of unlabelled speech -- a quantity comparable to what infants can be exposed to during language acquisition. Second, its functional hierarchy aligns with the cortical hierarchy of speech processing. Third, different training regimes reveal a functional specialization akin to the cortex: wav2vec 2.0 learns sound-generic, speech-specific and language-specific representations similar to those of the prefrontal and temporal cortices. Fourth, we confirm the similarity of this specialization with the behavior of 386 additional participants. These elements, resulting from the largest neuroimaging benchmark to date, show how self-supervised learning can account for a rich organization of speech processing in the brain, and thus delineate a path to identify the laws of language acquisition which shape the human brain.
\end{abstract}

\section{Introduction}
The performance of deep neural networks has taken off over the past decade. Algorithms trained on object classification, text translation, and speech recognition are starting to reach human-level performance \citep{xu2021self}. Furthermore, the \emph{representations} generated by these algorithms have repeatedly been shown to correlate with those of the brain \citep{kriegeskorte2015deep,yamins2016using,kietzmann2018deep,kell2019deep,cichy2019deep,toneva_interpreting_2019,millet2021inductive,caucheteux2022brains}, suggesting that these algorithms converge to brain-like computations. 

Such convergence, however, should not obscure the major differences that remain between these deep learning models and the brain. In particular, the above comparisons derive from models trained with 
(1) extraordinarily large amounts of data (40GB
for GPT-2 \citep{gpt2}, the equivalent of multiple lifetimes of reading), 
(2) supervised labels, which is rarely the case for humans (e.g. \citep{yamins2016using}),
(3) data in a textual rather than a raw sensory format, and/or
(4) considerable memory (e.g., language models typically have parallel access to thousands of context words to process text). These  differences highlight the pressing necessity to identify architectures and learning objectives which, subject to these four constraints, would be sufficient to account for both behavior and brain responses.

Here, we hypothesize that the latest self-supervised architectures trained on raw sensory data constitute promising candidates \citep{borgholt2022brief,bardes2021vicreg,Baevski2020wav2vec2A}. We focus on wav2vec 2.0 \citep{Baevski2020wav2vec2A}, an architecture that stacks convolutional and transformer layers to predict a quantization of the latent representations of speech waveforms. We train wav2vec 2.0 on 600\,h of \revised{effective} speech -- a quantity roughly comparable to what infants are exposed to during early language acquisition (speech only makes up a small fraction of infants' daily experience) \citep{dupoux2018cognitive,hart1992american,gilkerson2017mapping}.

\begin{figure}
  \centering
  \includegraphics[width=\textwidth]{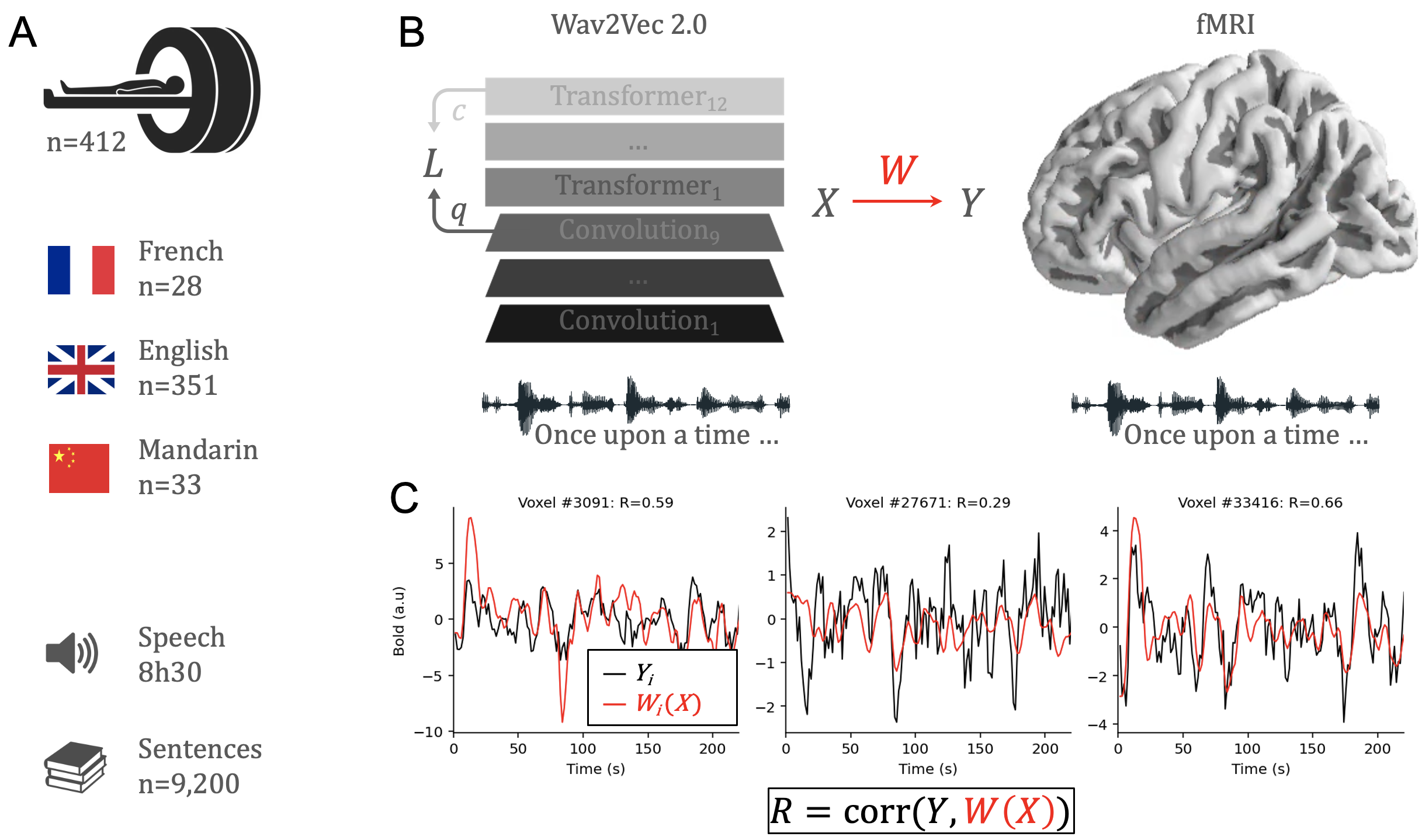}
  \caption{\textbf{Comparing speech representations in brains and deep neural networks.}
   \textbf{A.} We analyze the brain activity of 412 participants recorded with functional Magnetic Resonance Imaging (fMRI) while they passively listened to audio books in their native language (French, English or Mandarin). 
   \textbf{B.} After training wav2vec 2.0 \citep{Baevski2020wav2vec2A} with self-supervised learning ($L$) over 600\,h of unlabelled, effective speech, 
   we extract its activations in response to the audio books that were presented to the participants.
    We assess the similarity between the activations of the model $X$ and brain activity $Y$ with a standard encoding model $W$ \citep{nastase2020narratives} evaluated with a cross-validated Pearson correlation $R$. 
    \textbf{C.} Example\revised{s} of the true BOLD response (black) and the predicted BOLD response (red) estimated from a linear projection of the model's activations in \revised{three voxels randomly selected from the $10^{th}$ percentile of best voxels identified by the noise ceiling analysis for the first 200\,s of a representative story in the test set}.
   }
\label{fig-methods}
\end{figure}

We use standard encoding analyses  \citep{naselaris2011encoding,huth_natural_2016,yamins2016using,kell2018task} (Figure \ref{fig-methods}) to compare this model to the brains of 412 healthy volunteers (351 English speakers, 28 French speakers, and 33 Mandarin speakers) recorded with functional magnetic resonance imaging (fMRI) while they passively listened to approximately one hour of audio books in their native language \citep{nastase2020narratives,li2021petit} (8.5 hours of distinct audio materials in total).

To better understand the similarities between wav2vec 2.0 and the brain, we compare brain activity to each layer of this model, as well as to several variants, namely (1) a random (untrained) wav2vec 2.0 model, (2) a model trained on 600\,h of non-speech sounds, (3) a model trained on 600\,h of non-native speech  (for example, a model trained on English speech and mapped onto the brain responses to French-speaking participants), (4) a model trained on 600\,h of native speech  (for example, a model trained on English speech and mapped onto the brain responses to English participants), and (5) a model trained directly on speech-to-text (i.e., a supervised learning scheme) on the native language of the participants.

Our results provide four main contributions. First, self-supervised learning leads wav2vec 2.0 to learn latent representations of the speech waveform similar to those of the human brain. Second, the functional hierarchy of its transformer layers aligns with the cortical hierarchy of speech in the brain, and reveals the whole-brain organisation of speech processing with an unprecedented clarity. Third, the auditory-, speech-, and language-specific representations learned by the model converge to those of the human brain. Fourth, behavioral comparisons to 386 supplementary participants' results on a speech sound discrimination task confirm this common language specialization.

\begin{figure}[ht]
  \centering
  \includegraphics[width=\textwidth]{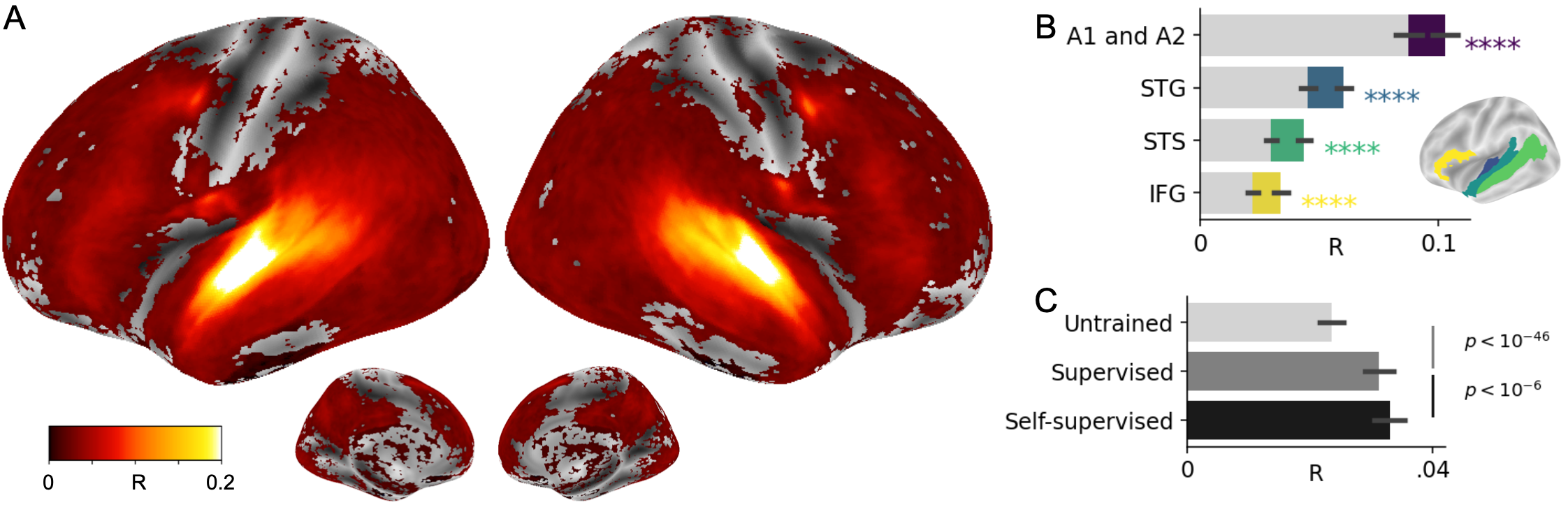}
  \caption{\textbf{Self-supervised learning suffices for wav2vec 2.0 to generate brain-like representations of speech.} 
  %
  \textbf{A.} 
  \revised{Brain score} ($R$) assessed for each subject and voxel independently, and here averaged across subjects for clarity. Only scores significantly above chance level, as assessed using a two-sided Wilcoxon test across subjects after correction for multiple comparison are color-coded ($p<10^{-10}$).
  \textbf{B.} $R$ scores for the same wav2vec2 model, averaged across subjects and voxels in four brain areas typically involved during speech processing (the primary and secondary auditory cortices, the superior temporal gyrus, the superior temporal sulcus, and the infero-frontal gyrus). In grey, the brain score obtained with a randomly initialized wav2vec 2.0 architecture. Error bars are the standard errors of the mean (SEM) across subjects. The stars indicate a significant difference between the random and trained model (all $p<10^{-4}$).
  \textbf{C.} $R$ scores of wav2vec 2.0 without training (top), trained with a supervised (middle) and self-supervised learning rule (bottom), on the same 600\,hours of effective speech. Scores are averaged across subjects and voxels and error bars are SEM across subjects.
  } 
\label{fig-prediction}
\end{figure}

\begin{figure} 
  \centering
  \includegraphics[width=0.9\textwidth]{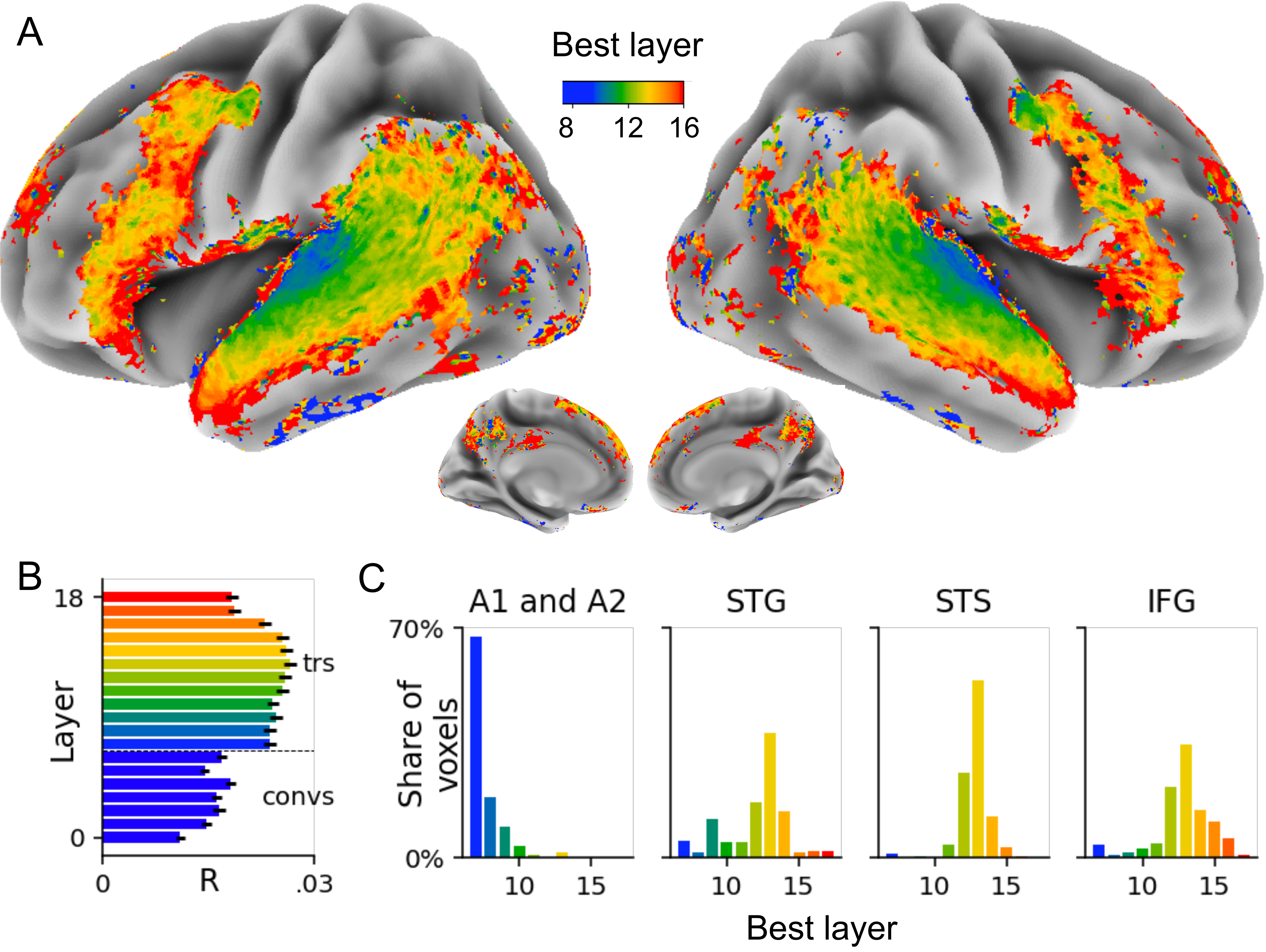}
  \caption{\textbf{The functional hierarchy of wav2vec 2.0 maps onto the speech hierarchy in the brain.} 
  \textbf{A.} We compute the $R$ score for each layer of wav2vec 2.0 separately and estimate, for each voxel, the layer with highest brain score on average across subjects. Only the voxels with significant brain scores are displayed ($p<10^{-18}$). While the first transformer layers (blue) map onto the low-level auditory cortices (A1 and A2), the deeper layers (orange and red) map onto brain regions associated with higher-level processes (e.g. STS and IFG).
  \textbf{B.} Layer-wise $R$ scores averaged across all voxels. Error bars are SEM across subjects. 
  \textbf{C.} Proportion of voxels with most predictive layer (x-axis) in four regions typically involved in speech processing. While most voxels in the primary cortex are best predicted by the first layers of the transformer, higher-level brain areas are best predicted by deeper layers. 
  }
\label{fig-hierarchy}
\end{figure}


\begin{figure}[ht]
  \centering
  \includegraphics[width=\textwidth]{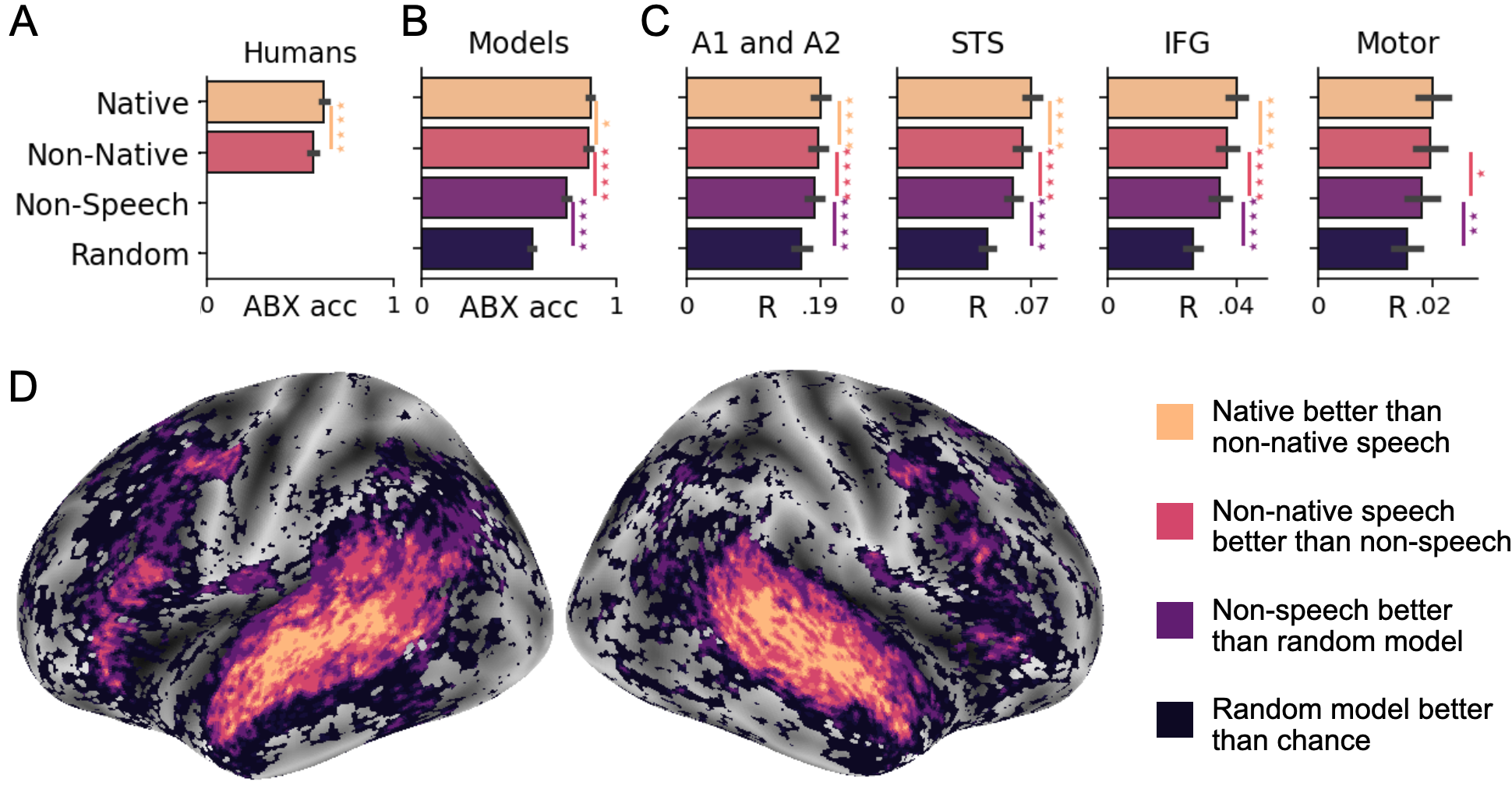}
\caption{\textbf{The specialization of wav2vec 2.0's representations follows and clarifies the acoustic, speech, and language regions in the brain.} \textbf{A.} We first evaluate humans' language specificity by quantifying their ability to perceive phonemes of their native or non-native languages (Section \ref{ABX}) in a ABX matching-to-sample task \citep{schatz2016abx} (higher is better). As expected, humans are better at matching phonemes of their native language. \textbf{B.} Then, we train four wav2vec 2.0 models with self-supervised learning on four datasets -- non-speech acoustic scenes, English, and French, and compute their ABX accuracy on the same speech datasets as humans. The `random' model is wav2vec 2.0 without any training. 
\textbf{C.} \revised{Brain score} ($R$) of each model (with an added model trained on Mandarin), averaged across voxels, in four regions of the brain (Section \ref{parcellation}). \textbf{D.} Acoustic, speech and language specificity for each voxel. For instance, one voxel is considered specific to the `native’ model if its native $R$ score is higher than its `non-native’ $R$ score ($p<.05$). Only the voxels with significant $R$ scores for the untrained model are displayed ($p<10^{-18}$). Error bars are the SEM across phone pairs in A and B, and across subjects in C. The stars indicates a significant difference between two conditions (Section \ref{stats}).
}
\label{fig-languages}
\end{figure}

\section{Methods}
\subsection{Models}

We train several variants of wav2vec 2.0 \citep{Baevski2020wav2vec2A} from scratch on different \revised{speech} datasets using two different learning objectives (a self-supervised and a supervised objective).

\subsubsection{Architecture}
Wav2vec 2.0 consists of three main modules. First, a feature encoder composed of seven blocks of temporal convolutions (output dimension 512) transforms the speech input $S$ (raw mono waveform at 16\,kHz) into a latent representation $z$ (output dimension of 512, frequency 49\,Hz, stride of 20\,ms between each frame, receptive field of 25\,ms). Second, a quantization module discretizes $z$ into $q$, a dictionary of discrete and latent representations of sounds. 
Third, $z$ is input to a ``context network'' consisting of 12 transformer blocks (model dimension 768, inner dimension 
3072, and 8 attention heads), which together yield a contextualized embedding $c$, of the same dimensionality of $q$. 

\subsubsection{Learning objective}

\paragraph{Self-supervised learning.} In this training paradigm, the model optimizes two losses. The first loss is contrastive and requires the model to predict the quantized representation $q$ of some masked input using $c$, from a finite set of quantized representations drawn from the input sample. The second loss ensures that the quantized representations are diverse. See Section \ref{ap:loss_sup} and \citep{Baevski2020wav2vec2A} for details.

\paragraph{Supervised learning.} In this training paradigm, the quantization module is discarded and a linear layer mapping $c$ to phonemes is added at the end of the pipeline. The model is randomly initialized and all layers (including the feature encoder) are trained using a Connectionist Temporal Classification (CTC) \citep{graves2012connectionist} loss to perform phone recognition.
For both training paradigms, we extract the activations of each layer from both the feature encoder (outputting $z$) and the context network (outputting $c$). We extract the representations of the convolutional and transformer blocks using an input window of 10\,s of raw waveform (stride = 5\,s).

\subsubsection{Training}
\label{sec-trainingimplem}
\paragraph{Datasets.} We successively train different wav2vec 2.0 models using each of four datasets: (i) the French and (ii) English CommonVoice corpora \citep{ardila2020common}, (iii) the MAGICDATA Mandarin Chinese Read Speech Corpus \citep{magicdata}, and (iv) a non-speech subset of the Audioset dataset \citep{gemmeke2017audio}, which contains recordings of various acoustic scenes.

\paragraph{Preprocessing.} All the audio datasets were randomly subsampled to have an approximate size of 600 hours, downsampled to 16\,kHz and converted to mono with the Sox software\footnote{\url{http://sox.sourceforge.net/}}. We randomly split the datasets into a training (80\%), a validation (10\%) and a test set (10\%). 
The audio recordings we use from the Audioset dataset are filtered so that they do not contain speech or any sounds produced by humans, such as laughter or singing. 
For the speech datasets, we also use their corresponding annotations (in the supervised settings). We phonemize these annotations using eSpeakNG\footnote{\url{https://github.com/espeak-ng/espeak-ng}}. The number of different phoneme symbols in these annotations is similar for French (32), English (39), and Mandarin Chinese (33).


\paragraph{Implementation.} We train all of our models using the fairseq implementation of wav2vec 2.0\footnote{\url{ https://github.com/pytorch/fairseq/tree/main/examples/wav2vec}} using default hyperparameters.  We also analyze a model whose parameters were randomly initialised (``untrained'' model).

We use self-supervised learning to train four models: three on the speech datasets (French, English, and Mandarin) and one on the acoustic scenes dataset. In each case, the training was performed using the same configuration file (namely, the base configuration provided in the fairseq repository for pretraining wav2vec 2.0 on LibriSpeech \citep{panayotov2015librispeech}). We train the models for 400k updates and select the ones with the best validation loss.

We also use the supervised training paradigm to train three models, on the French, English, and Mandarin datasets, respectively. Each training was performed using the same configuration file, which was identical to the configuration provided in the fairseq repository for  fine-tuning wav2vec 2.0 on the 960 hour Voxpopuli corpus \citep{wang2021voxpopuli}, except that parameters were not frozen (\texttt{freeze\_finetune\_updates$=0$}) and learning was performed on all parameters of the models using the CTC loss (\texttt{feature\_grad\_mult$=0.1$}). We train the models for 400k updates and we use the ones with the best word error rate (WER) on the validation set. The French model obtains 13.9 WER, the English model 28.6 WER, and the Mandarin model 4.6 WER, on their respective test sets.


\subsection{Functional MRI}
\label{sec-fmridata}
We analyse a composite set of fMRI recordings aggregated from the \textit{Little Prince} \citep{li2021petit} and the \textit{Narratives} public datasets \citep{nastase2020narratives}.


\paragraph{Narratives.} This dataset\footnote{\url{https://openneuro.org/datasets/ds002345}} contains the fMRI recordings of 345 native English-speaking participants listening to English narratives (4.6 hours of unique audio in total). \revised{The participants listened to different stories varying from 7 to 98\,min (mean\,=\,26\,min)}. Following \citep{nastase2020narratives}, we (1) focus on fifteen representative stories and ignore the narratives that have been modified by scrambling and (2) exclude eight participants because of noisy recordings. Overall, this selection results in a curated dataset of 303 participants listening to fifteen stories ranging from 3\,min to 56\,min, for a total of 4 hours of unique audio (36,018 words from a vocabulary of 4,004 unique words).

\paragraph{The Little Prince.} This dataset\footnote{\url{https://openneuro.org/datasets/ds003643/versions/1.0.4}} contains fMRI recordings of 48 English native speakers, 33 Mandarin native speakers, and 28 French native speakers listening to \textit{The Little Prince} in their respective native language.
The experiment itself was divided into nine runs of approximately 10\,min of passive listening. 
%
For each language condition, the story was read by a single native speaker. 
The English, Mandarin, and French audiobooks last 94, 90 and 97 minutes respectively. 

\paragraph{Preprocessing.}
For Narratives, we did not perform additional preprocessing: we use the public preprocessing of the dataset already projected on the surface space (``fsaverage6'') without spatial smoothing (labelled ``afni-nosmooth'' in the data repository). In contrast, the \textit{Little Prince} dataset is only provided in a volumetric space. Consequently, for each language condition separately, we subselected the cortical voxels by computing a brain mask using the average of all participants' fMRI data realigned onto a common template brain via Freesurfer \citep{freesurfer}. These voxels are then projected onto a brain surface using nilearn's {\texttt{vol\_to\_surf}} function with defaults parameters \citep{nilearn} and a `fsaverage6` template surface \citep{freesurfer}. 
For both \textit{Narratives} and \textit{The Little Prince}, fMRI signals are normalized across the time dimension to have a mean of 0 and a variance of 1, for each participant, surface voxel and session independently.

\paragraph{Brain parcellation.} \label{parcellation}
For the purposes of certain analyses, 
we group the fMRI voxels into regions of interest using the Destrieux Atlas \citep{destrieux_automatic_2010}. This parcellation results in 75 brain regions in each hemisphere. For simplicity, we label the regions as follows: A1 and A2 represents Heschl gyrus, which is the anatomical location of the primary and secondary auditory cortices, STG and STS are the superior temporal gyrus and sulcus, and IFG is the inferior frontal gyrus.

\subsection{Brain score (R)}\label{np_score}

To quantify the similarity between the network's activations $X$ and the brain recordings $Y$, we use a standard linear encoding model \citep{huth_natural_2016,yamins2016using}. For each subject, we split the data into train and test sets using a five-fold cross-validation setting. For each train split, a linear mapping $W$ is fitted to predict the brain response $Y_{\mathrm{train}}$ given $X_{\mathrm{train}}$. $W$ combines a temporal alignment function with fixed weight, and a trained penalized linear regression.

\paragraph{Temporal alignment.} The sampling frequency of the model's activations (between 49 and 200\,Hz) differs from the sampling frequency of fMRI BOLD signals (0.5\,Hz). Furthermore, the BOLD signals have delayed responses spanning over several seconds. Thus, we first convolve the model activations  with a standard hemodynamic response function (HRF) using nistats \citep{nilearn} \texttt{compute\_regressor} function with the `glover' model and default parameters. This results in the convolved activations $X'_{\mathrm{train}}$ with the same sampling frequency as the fMRI $Y_{\mathrm{train}}$ (see \ref{ap:preproc}). 


\paragraph{Penalised linear regression.} Once temporally aligned, we fit an $\ell_2$-penalised linear regression that predicts the brain signals $Y_{\mathrm{train}}$ given the activations $X_{\mathrm{train}}$. We use the \texttt{RidgeCV} function from scikit-learn \citep{pedregosa2011scikit}, with the penalization hyperparameter $\lambda$ varying between 10 and $10^8$ (20 values scaled logarithmically) chosen independently for each dimension with a nested cross-validation over the training set (see \ref{ap:ridge}).

\paragraph{Evaluation.} We evaluate the linear mapping $W$ on the held out sets by measuring Pearson's correlation between predicted and actual brain responses:
%
    $R = \mathrm{corr}\big(Y_{\mathrm{test}}, W \cdot X_{\mathrm{test}}\big)$. 
%
Finally, we average the correlation scores across test splits to obtain the final ``brain score''. 
\label{circle_mean}
To report the average layer $k^*$ with the highest brain score for each voxel (Figure \ref{fig-hierarchy}), while being robust to regression-to-the-mean biases, we first find the best layer $k_s$ for each participant $s$ and each voxel independently and then compute a circular mean across the $N=412$ participants and the $K=19$ layers:
$
    k^* = \mathrm{angle}\left(\frac{1}{N}\sum_{s=1}^N \exp\left(\frac{2i\pi k_s}{K+1}\right)\right)
$

\paragraph{Statistics.}\label{stats}
We assess the reliability of brain scores with second-level analyses across participants thanks to a Wilcoxon signed-rank test across participants. Thus, the resulting p-values are not affected by fMRI auto-correlation within participants. We perform statistical correction for multiple comparisons with Benjamini–Hochberg False Discovery Rate (FDR) across voxels \citep{benjamini2010discovering}.

\subsection{Behavioral experiment} \label{ABX}
\revised{To compare the phonetic representations of our models to those of humans, we compare the forced-choice discrimination judgements of online participants\footnote{\url{https://docs.cognitive-ml.fr/perceptimatic/}} to an analogous method applied to wav2vec 2.0 \citep{schatz2016abx}. 
Specifically, for each triplet of sound ``ABX'', participants judged whether the stimulus X was more similar to A or B. Analogously, we computed the Euclidean distance in the most discriminative layer of wav2vec 2.0 (here transformer layer 5) to determine whether X was closer to A or B. Additional data, analyses and model-human comparison can be found in \citep{millet2022self}.}
%
%
We focus on the French and English stimuli, which represent $\approx$ 6,000 ABX triplets (testing 508 English and 524 French phone pairs), with 386 participants in total (193 from each language group). 

In Figure \ref{fig-languages}-A, we report the ABX accuracy of English- and French-speaking participants in both their native and non-native language (either English or French). We first average results per phone pair, and then average over phone pairs to obtain the ABX discrimination accuracy. Similarly, in Figure \ref{fig-languages}-B, we compute the ABX accuracy of our wav2vec 2.0 models on the same evaluation sets as the participants, using the parameters described in \citep{millet2022self}. English and French models are evaluated on the same (`native') or different (`non-native') language stimuli as their training. The random and non-speech models are evaluated on both French and English speech stimuli.



\section{Results}
\label{sec:results}

\paragraph{Wav2vec 2.0 maps onto brain responses to speech.}
We estimate whether the activations of wav2vec 2.0 models linearly map onto the human brain activity of 412 individuals listening to audio books in the fMRI scanner.
%
%
%
For this, we first independently train three models with 600\,h of French, English, or Mandarin, respectively, and 
compute the \revised{brain scores} ($R$) with the corresponding participants. Specifically, we (1) convolve the activations ($X$) of the model with a hemodynamic response function (HRF), (2) train a $\ell_2$-penalized linear regression on a training split to map them to brain activity $Y$, and (3) compute the Pearson correlation coefficient between (i) the true fMRI activity and (ii) the predicted activations on a test split.
%
The models' activations significantly predict brain activity in nearly all cortical areas, reaching the highest $R$ scores in the primary and secondary auditory cortices (Figure \ref{fig-prediction}-A\,B). These scores are significantly higher than those obtained with a randomly initialised model ($p<10^{-50}$ on average across voxels), and this comparison is robust across language groups (all $p<10^{-5}$).
%




\paragraph{Comparison of self-supervised to supervised models.}
Does self-supervision reach representations that are as brain-like as those obtained with supervised learning?
To address this issue, we trained wav2vec 2.0 with an alternative, supervised objective, namely, predicting phonetic annotations from the same 600 hours of \revised{effective} speech sounds. We then implemented the $R$ score analyses described above. The results show that self-supervised learning in fact leads to modestly but significantly better $R$ scores than supervised learning (Figure \ref{fig-prediction}-C): $\Delta R = 0.002, p<10^6$. 



\paragraph{The hierarchy of wav2vec 2.0 maps onto the hierarchy of the cortex.
}
To compare the speech hierarchy in the brain with the functional hierarchy learned by wav2vec 2.0, we evaluate the $R$ score of each layer of the model (Figure \ref{fig-hierarchy}). 
First, we observe that convolutional layers are less predictive than transformer layers. Second, within the transformers, the hierarchy of representations aligns with the expected cortical hierarchy \citep{hickok2007cortical}: while low-level areas (A1, A2) are best predicted by the first transformer layers, higher level areas (IFG, STS) are best predicted by deeper layers. Remarkably, this hierarchy extends to supplementary motor and motor areas in both hemispheres (Figure \ref{fig-hierarchy}-A). 

\paragraph{Language specificity in phone discrimination tasks.}

The acoustic features underlying speech (fricatives, vowels, and so on) may also characterize non-speech sounds (the sound of tree leaves in the wind, of a stone falling, and so on). Does the model show commonalities merely with general auditory processing in the brain, or does it capture speech-specific processing? If so, does it show commonalities with brain representations that are specific to the native language of the participants, or merely to general speech processing?
%
We first evaluate the specialization of humans' perception to their native language using an ABX behavioral task (Section \ref{ABX}). Specifically, we compare 386 French and English participants on their ability to distinguish native and non-native phones. 
As expected \citep{bohn2017cross,kuhl2005early}, participants were better at discriminating native sounds than non-native ones (across phone pairs: $p<10^{-18}$, Figure \ref{fig-languages}-A). 
Second, applying the same test to our self-supervised French and English models shows that, like humans, 
models best discriminate sounds from their `native' language (i.e., the French model better distinguishes French stimuli than English ones, across phone pairs, and vice versa: $p<0.05$). 
\revised{
Interestingly, the ABX accuracy of the model is significantly higher than participants’. This quantitative difference may be partially explained by the fact that participants – and online participants in particular – undergo fluctuating attention, and adopt strategies which can negatively impact performance \citep{humphreys_acquisition_1939}. 
}
Finally, as expected, the random and acoustic models 
obtain the worst ABX accuracy. 
Overall, These results confirm that 600\,h of self-supervised learning on \revised{effective} speech suffices for wav2vec 2.0 to learn language-specific representations (Figure \ref{fig-languages}-B).

\paragraph{Wav2vec 2.0 and the brain learn language specific representations.}

Next, we compare the \revised{brain scores} of random, non-speech, non-native and native models (Figure \ref{fig-languages}-C\,D). 
%
First, our results show that the non-speech model attains higher $R$ scores than the random model (on average across voxels, $\Delta R = 0.006$, $p=10^{-31}$) confirming the importance of learning to generate brain-like representations. Second, non-native models attain higher $R$ scores than the non-speech model ($\Delta R=0.002, p=10^{-9}$), confirming that wav2vec 2.0 learns speech-specific representations of sounds when trained on speech. Finally, the native model attains higher $R$ scores than non-native models ($\Delta R=0.002, p=10^{-15}$). 
%
%
%



\section{Discussion}
\label{sec-discussion}

Human infants acquire language with little to no supervision: A few hundred hours of speech suffices for their young brain to learn to discretize phonemes, segment morphemes, and assemble words in the language(s) of their social group \citep{dupoux2018cognitive,gilkerson2017mapping}. 
\revised{However, the learning principle that allows this unique feat remains, to date, unknown.

Here, we test whether self-supervised learning applied to a limited amount of speech effectively accounts for the organization of speech processing in the human brain as measured with fMRI. For this, }we train several variants of wav2vec 2.0 \citep{Baevski2020wav2vec2A} with three curated datasets of French, English, and Mandarin, and compare their activations to those of a large group of French, English, and Mandarin speakers recorded with fMRI while passively listening to audio stories.
Our results show that this self-supervised model learns (i) representations that linearly map onto a remarkably distributed set of cortical regions (Figure \ref{fig-prediction}), (ii) a \revised{computational} hierarchy that aligns with the cortical hierarchy (Figure \ref{fig-hierarchy}), and (iii) features specific to the language of the participants (Figure~\ref{fig-languages}). 


\paragraph{Towards a biologically-plausible learning principle.} These results extend recent findings on the similarities between the brain and a variety of deep learning models trained with biologically-implausible objectives and data.
First, fMRI \citep{kell2018task,millet2021inductive,thompson2021training}, electroencephalography \citep{huang2018connecting}, and multi- or single-unit responses to sounds  \citep{koumura2019cascaded,begus2022encoding} have been shown to be linearly predicted by the activations of deep convolutional networks trained on \emph{supervised} auditory tasks. For example, \citep{millet2021inductive} showed that a supervised \revised{speech-to-text} model better accounted for brain responses to speech in 102 individuals when it was trained on speech recognition rather than auditory scene classification. Similarly, \citep{kell2018task} showed that eight participants listening to brief speech and non-speech sounds demonstrated fMRI responses in the temporal lobe that aligned with those of a deep convolutional neural network trained on a binary auditory classification task. 
Our results, based on up to 50 times more fMRI recordings of the entire cortex show that such representational similarities hold with a self-supervised objective \citep{lerner_topographic_2011, berezutskaya2017neural,caucheteux_model-based_2021,caucheteux_long-range_2021}.
%
Second, a growing series of MEG \citep{toneva_interpreting_2019,caucheteux2022brains}, fMRI \citep{mitchell2008predicting,qian2016bridging,pereira2018toward,schwartz2019inducing,antonello2021low,jain2018incorporating} and electro-physiology studies \citep{schrimpf2021neural,goldstein2022shared} showed that text-based language models trained on very large corpora generate brain-like representations too. 
While these results suggest elements of convergence between language models and the brain \citep{caucheteux2022brains}, they also remain biologically implausible: not only are these algorithms pre-equipped with abstract linguistic units such as characters and words,  but they are trained on corpora that no one would ever be able to read in their lifetime. In contrast, wav2vec 2.0 is here trained with a reasonable amount of raw speech waveforms \citep{hart1992american,gilkerson2017mapping,dupoux2018cognitive}. \revised{The functional similarity between wav2vec 2.0 and the brain thus opens the way to clarify how humans learn to process speech.}

\paragraph{The emergence of a brain-like hierarchy of speech processing.}
The present study reveals the hierarchical organization of speech processing with remarkable clarity.
First, the functional hierarchy learnt by wav2vec 2.0 is aligned with the anatomy: \emph{e.g.} the superior temporal sulcus and the temporal pole are known to project to the ventral and dorsal part of the inferofrontal gyrus, respectively \citep{petkov_different_2015}.
Second, the identification of functional gradients within the prefrontal cortex, and down to the motor areas typically associated with larynx and mouth control \citep{dichter2018control} reinforces the relevance of motor processes to speech perception \citep{kellis2010decoding,mugler2014direct,shamma2021learning}. 
Finally, the existence of multiple levels of representations around the inferofrontal cortex
is consistent with the idea that Broca's area may be responsible for merging linguistic units 
\citep{chomsky2000linguistics,friederici1999neurobiology,hagoort2005broca,poeppel2012towards}. It should be noted, however, that our results aggregate a large cohort of individuals which could mask a more modular organization at the individual level.

\revised{
\paragraph{Interpreting the neural representations of speech. } 
Interpreting neural representations is a notoriously difficult challenge to both AI and neuroscience. Here, we first investigate language specificity and show that the neural representations specific to the native models are primarily represented in the superior temporal sulcus and middle temporal gyrus  (Figure \ref{fig-languages}D): areas known to represent phonetic features \citep{mesgarani_phonetic_2014}. 
%
%
%
However, these effect are relatively modest (Figure \ref{fig-languages}): the random model and the non-speech model reach, in STS and STG, 67\% and 87\% of the \revised{brain scores} obtained by the ``native'' model, respectively. While this high baseline initially surprised us, this phenomenon could be explained by the fact that the auditory cortex is continuously bombarded by -- and should thus be tuned to -- non-speech input. 
}
\revised{Second, our probing analyses show that the models trained with self-supervised learning learn relevant acoustic and linguistic representations (Supplementary Figure \ref{fig-probe}). This result, consistent with \citet{vaidya2022self} and \citet{stephenson2019untangling}, suggests that the difference of \revised{brain scores} observed between the random, non-native and native models (Figure \ref{fig-languages}) may be partly driven by the corresponding spectro-temporal, phonetic, word and sentence-level representations, respectively. These elements of interpretation remain, however, scarce, and a systematic interpretation of the representations shared between wav2vec 2.0 and the brain remains necessary.  
}

\revised{ 
\paragraph{Scope of the study.} It is important to stress that the scope of the present study could be broadened in several ways. First, our study focuses on adult speakers, whose cultural and educational background is not representative of the population \citep{henrich_weirdest_2010}. Second, we focus on the passive listening of three languages. Third, we focus on one self-supervised learning architecture \citep{Baevski2020wav2vec2A}, and its functional alignment with fMRI, whose temporal resolution is notoriously limited. Generalizing the present approach to more languages \citep{malik-moraleda_investigation_2022}, a larger spectrum of children and adult participants recorded with a variety of electrophysiological and neuroimaging devices will thus be essential to confirm, precise, and/or mitigate the present findings.}

\revised{ 
\paragraph{The remaining gap between brain and speech models.} 
Several major gaps can be evidenced between wav2vec 2.0 and the brain. First, the transformer layers are not temporally constrained: each layer can access all elements within the contextual window. This differs from the necessarily recurrent nature of processing in the brain. Second, wav2vec 2.0 behaves differently to humans in specific tasks. In particular, it is overly-sensitive to band-pass filtering, non-robustly exploit fine temporal structures \citep{weerts2021psychometrics} and fails to display the expected categorical responses \citep{millet-etal-2021-predicting}. Third, recent studies show that wav2vec 2.0 encodes significantly less semantic information than text-based models \citep{pasad2021layer,vaidya2022self}. While our analyses suggest that learning allows wav2vec 2.0 to capture some lexical features in its deep layers (Figure \ref{fig-probe}, Table \ref{tab:probe}), it remains unclear whether these layers also capture complex syntactic structures, such as recursive syntactic trees \citep{lakretz_can_2021, caucheteux_disentangling_2021}. We speculate that these limitations may be due to the time scales of wav2vec 2.0 which, unlike humans, learns very short-term representations of speech.
In any case, these differences likely explain why the \revised{brain scores} of wav2vec 2.0 remain substantially lower than our noise-ceiling (19\% on average, and up to 74\% in Heschl’s gyrus and sulcus, Table \ref{tab:np_noiseceil}, Figure \ref{fig-noiseceil}). 

}

Overall, the complexity of the human brain is often thought to be incompatible with a simple theory: ``Even if there were enough data available about the contents of each brain area, there probably would not be a ready set of equations to describe them, their relationships, and the ways they change over time''  \citep{gallant2013reading}. By showing how the equations of self-supervised learning give rise to brain-like processes, this work contributes to challenge this view.


\section*{Acknowledgments}
This project was funded, in part, by the Bettencourt-Schueller Foundation, the Philippe Foundation, and FrontCog grant ANR-17-EURE-0017 to JRK for his work at PSL.

\bibliography{50_biblio}

\begin{thebibliography}{75}
\providecommand{\natexlab}[1]{#1}
\providecommand{\url}[1]{\texttt{#1}}
\expandafter\ifx\csname urlstyle\endcsname\relax
  \providecommand{\doi}[1]{doi: #1}\else
  \providecommand{\doi}{doi: \begingroup \urlstyle{rm}\Url}\fi

\bibitem[Abraham et~al.(2014)Abraham, Pedregosa, Eickenberg, Gervais, Mueller,
  Kossaifi, Gramfort, Thirion, and Varoquaux]{nilearn}
Alexandre Abraham, Fabian Pedregosa, Michael Eickenberg, Philippe Gervais,
  Andreas Mueller, Jean Kossaifi, Alexandre Gramfort, Bertrand Thirion, and
  Ga{\"e}l Varoquaux.
\newblock Machine learning for neuroimaging with scikit-learn.
\newblock \emph{Frontiers in neuroinformatics}, 8:\penalty0 14, 2014.

\bibitem[Antonello et~al.(2021)Antonello, Turek, Vo, and
  Huth]{antonello2021low}
Richard Antonello, Javier~S Turek, Vy~Vo, and Alexander Huth.
\newblock Low-dimensional structure in the space of language representations is
  reflected in brain responses.
\newblock \emph{Advances in Neural Information Processing Systems}, 34, 2021.

\bibitem[Ardila et~al.(2020)Ardila, Branson, Davis, Kohler, Meyer, Henretty,
  Morais, Saunders, Tyers, and Weber]{ardila2020common}
Rosana Ardila, Megan Branson, Kelly Davis, Michael Kohler, Josh Meyer, Michael
  Henretty, Reuben Morais, Lindsay Saunders, Francis~M Tyers, and Gregor Weber.
\newblock Common voice: A massively-multilingual speech corpus.
\newblock In \emph{LREC}, 2020.

\bibitem[Baevski et~al.(2020)Baevski, Zhou, rahman Mohamed, and
  Auli]{Baevski2020wav2vec2A}
Alexei Baevski, H.~Zhou, Abdel rahman Mohamed, and Michael Auli.
\newblock wav2vec 2.0: A framework for self-supervised learning of speech
  representations.
\newblock \emph{ArXiv}, abs/2006.11477, 2020.

\bibitem[Bardes et~al.(2021)Bardes, Ponce, and LeCun]{bardes2021vicreg}
Adrien Bardes, Jean Ponce, and Yann LeCun.
\newblock Vicreg: Variance-invariance-covariance regularization for
  self-supervised learning.
\newblock \emph{arXiv preprint arXiv:2105.04906}, 2021.

\bibitem[Begus et~al.(2022)Begus, Zhou, and Zhao]{begus2022encoding}
Gasper Begus, Alan Zhou, and Christina Zhao.
\newblock Encoding of speech in convolutional layers and the brain stem based
  on language experience.
\newblock \emph{bioRxiv}, 2022.

\bibitem[Benjamini(2010)]{benjamini2010discovering}
Yoav Benjamini.
\newblock Discovering the false discovery rate.
\newblock \emph{Journal of the Royal Statistical Society: series B (statistical
  methodology)}, 72\penalty0 (4):\penalty0 405--416, 2010.

\bibitem[Berezutskaya et~al.(2017)Berezutskaya, Freudenburg,
  G{\"u}{\c{c}}l{\"u}, van Gerven, and Ramsey]{berezutskaya2017neural}
Julia Berezutskaya, Zachary~V Freudenburg, Umut G{\"u}{\c{c}}l{\"u}, Marcel~AJ
  van Gerven, and Nick~F Ramsey.
\newblock Neural tuning to low-level features of speech throughout the
  perisylvian cortex.
\newblock \emph{Journal of Neuroscience}, 37\penalty0 (33):\penalty0
  7906--7920, 2017.

\bibitem[Bohn(2017)]{bohn2017cross}
Ocke-Schwen Bohn.
\newblock Cross-language and second language speech perception.
\newblock \emph{The handbook of psycholinguistics}, pages 213--239, 2017.

\bibitem[Borgholt et~al.(2022)Borgholt, Havtorn, Edin, Maal{\o}e, and
  Igel]{borgholt2022brief}
Lasse Borgholt, Jakob~Drachmann Havtorn, Joakim Edin, Lars Maal{\o}e, and
  Christian Igel.
\newblock A brief overview of unsupervised neural speech representation
  learning.
\newblock \emph{arXiv preprint arXiv:2203.01829}, 2022.

\bibitem[Caucheteux and King(2022)]{caucheteux2022brains}
Charlotte Caucheteux and Jean-R{\'e}mi King.
\newblock Brains and algorithms partially converge in natural language
  processing.
\newblock \emph{Communications Biology}, 5\penalty0 (1):\penalty0 1--10, 2022.

\bibitem[Caucheteux et~al.(2021{\natexlab{a}})Caucheteux, Gramfort, and
  King]{caucheteux_disentangling_2021}
Charlotte Caucheteux, Alexandre Gramfort, and Jean-Remi King.
\newblock Disentangling syntax and semantics in the brain with deep networks.
\newblock In \emph{Proceedings of the 38th {International} {Conference} on
  {Machine} {Learning}}, pages 1336--1348. PMLR, July 2021{\natexlab{a}}.
\newblock ISSN: 2640-3498.

\bibitem[Caucheteux et~al.(2021{\natexlab{b}})Caucheteux, Gramfort, and
  King]{caucheteux_long-range_2021}
Charlotte Caucheteux, Alexandre Gramfort, and Jean-Remi King.
\newblock Long-range and hierarchical language predictions in brains and
  algorithms.
\newblock \emph{arXiv:2111.14232 [cs, q-bio]}, November 2021{\natexlab{b}}.
\newblock arXiv: 2111.14232.

\bibitem[Caucheteux et~al.(2021{\natexlab{c}})Caucheteux, Gramfort, and
  King]{caucheteux_model-based_2021}
Charlotte Caucheteux, Alexandre Gramfort, and Jean-Remi King.
\newblock Model-based analysis of brain activity reveals the hierarchy of
  language in 305 subjects.
\newblock In \emph{Findings of the {Association} for {Computational}
  {Linguistics}: {EMNLP} 2021}, pages 3635--3644, Punta Cana, Dominican
  Republic, November 2021{\natexlab{c}}. Association for Computational
  Linguistics.

\bibitem[Caucheteux et~al.(2022)Caucheteux, Gramfort, and
  King]{caucheteux_gpt-2s_2021}
Charlotte Caucheteux, Alexandre Gramfort, and Jean-Rémi King.
\newblock Deep language algorithms predict semantic comprehension from brain
  activity.
\newblock \emph{Scientific Reports}, 12\penalty0 (1):\penalty0 16327, September
  2022.
\newblock ISSN 2045-2322.
\newblock \doi{10.1038/s41598-022-20460-9}.
\newblock Number: 1 Publisher: Nature Publishing Group.

\bibitem[Chomsky(2000)]{chomsky2000linguistics}
Noam Chomsky.
\newblock Linguistics and brain science.
\newblock \emph{Image, language, brain}, pages 13--28, 2000.

\bibitem[Cichy and Kaiser(2019)]{cichy2019deep}
Radoslaw~M Cichy and Daniel Kaiser.
\newblock Deep neural networks as scientific models.
\newblock \emph{Trends in cognitive sciences}, 23\penalty0 (4):\penalty0
  305--317, 2019.

\bibitem[Co.(2019)]{magicdata}
Beijing Magic Data~Technology Co.
\newblock Magic data chinese mandarin conversational speech.
\newblock
  \url{http://www.imagicdatatech.com/index.php/home/dataopensource/data_info/id/101},
  2019.

\bibitem[Destrieux et~al.(2010)Destrieux, Fischl, Dale, and
  Halgren]{destrieux_automatic_2010}
Christophe Destrieux, Bruce Fischl, Anders Dale, and Eric Halgren.
\newblock Automatic parcellation of human cortical gyri and sulci using
  standard anatomical nomenclature.
\newblock \emph{NeuroImage}, 53\penalty0 (1):\penalty0 1--15, October 2010.
\newblock ISSN 1053-8119.
\newblock \doi{10.1016/j.neuroimage.2010.06.010}.

\bibitem[Dichter et~al.(2018)Dichter, Breshears, Leonard, and
  Chang]{dichter2018control}
Benjamin~K Dichter, Jonathan~D Breshears, Matthew~K Leonard, and Edward~F
  Chang.
\newblock The control of vocal pitch in human laryngeal motor cortex.
\newblock \emph{Cell}, 174\penalty0 (1):\penalty0 21--31, 2018.

\bibitem[Dupoux(2018)]{dupoux2018cognitive}
Emmanuel Dupoux.
\newblock Cognitive science in the era of artificial intelligence: A roadmap
  for reverse-engineering the infant language-learner.
\newblock \emph{Cognition}, 173:\penalty0 43--59, 2018.

\bibitem[Fischl(2012)]{freesurfer}
Bruce Fischl.
\newblock Freesurfer.
\newblock \emph{Neuroimage}, 62\penalty0 (2):\penalty0 774--781, 2012.

\bibitem[Friederici(1999)]{friederici1999neurobiology}
Angela~D Friederici.
\newblock The neurobiology of language comprehension.
\newblock In \emph{Language comprehension: A biological perspective}, pages
  265--304. Springer, 1999.

\bibitem[Gallant(2013)]{gallant2013reading}
Jack Gallant.
\newblock in 'reading minds'.
\newblock \emph{Nature}, 502\penalty0 (7472):\penalty0 428, 2013.

\bibitem[Gemmeke et~al.(2017)Gemmeke, Ellis, Freedman, Jansen, Lawrence, Moore,
  Plakal, and Ritter]{gemmeke2017audio}
Jort~F Gemmeke, Daniel~PW Ellis, Dylan Freedman, Aren Jansen, Wade Lawrence,
  R~Channing Moore, Manoj Plakal, and Marvin Ritter.
\newblock Audio set: An ontology and human-labeled dataset for audio events.
\newblock In \emph{2017 IEEE International Conference on Acoustics, Speech and
  Signal Processing (ICASSP)}, pages 776--780. IEEE, 2017.

\bibitem[Gilkerson et~al.(2017)Gilkerson, Richards, Warren, Montgomery,
  Greenwood, Kimbrough~Oller, Hansen, and Paul]{gilkerson2017mapping}
Jill Gilkerson, Jeffrey~A Richards, Steven~F Warren, Judith~K Montgomery,
  Charles~R Greenwood, D~Kimbrough~Oller, John~HL Hansen, and Terrance~D Paul.
\newblock Mapping the early language environment using all-day recordings and
  automated analysis.
\newblock \emph{American journal of speech-language pathology}, 26\penalty0
  (2):\penalty0 248--265, 2017.

\bibitem[Goldstein et~al.(2022)Goldstein, Zada, Buchnik, Schain, Price, Aubrey,
  Nastase, Feder, Emanuel, Cohen, et~al.]{goldstein2022shared}
Ariel Goldstein, Zaid Zada, Eliav Buchnik, Mariano Schain, Amy Price, Bobbi
  Aubrey, Samuel~A Nastase, Amir Feder, Dotan Emanuel, Alon Cohen, et~al.
\newblock Shared computational principles for language processing in humans and
  deep language models.
\newblock \emph{Nature neuroscience}, 25\penalty0 (3):\penalty0 369--380, 2022.

\bibitem[Graves(2012)]{graves2012connectionist}
Alex Graves.
\newblock Connectionist temporal classification.
\newblock In \emph{Supervised Sequence Labelling with Recurrent Neural
  Networks}, pages 61--93. Springer, 2012.

\bibitem[Hagoort(2005)]{hagoort2005broca}
Peter Hagoort.
\newblock On broca, brain, and binding: a new framework.
\newblock \emph{Trends in cognitive sciences}, 9\penalty0 (9):\penalty0
  416--423, 2005.

\bibitem[Hart and Risley(1992)]{hart1992american}
Betty Hart and Todd~R Risley.
\newblock American parenting of language-learning children: Persisting
  differences in family-child interactions observed in natural home
  environments.
\newblock \emph{Developmental psychology}, 28\penalty0 (6):\penalty0 1096,
  1992.

\bibitem[Henrich et~al.(2010)Henrich, Heine, and
  Norenzayan]{henrich_weirdest_2010}
Joseph Henrich, Steven~J. Heine, and Ara Norenzayan.
\newblock The weirdest people in the world?
\newblock \emph{Behavioral and Brain Sciences}, 33\penalty0 (2-3):\penalty0
  61--83, June 2010.
\newblock ISSN 0140-525X, 1469-1825.
\newblock \doi{10.1017/S0140525X0999152X}.

\bibitem[Hickok and Poeppel(2007)]{hickok2007cortical}
Gregory Hickok and David Poeppel.
\newblock The cortical organization of speech processing.
\newblock \emph{Nature reviews neuroscience}, 8\penalty0 (5):\penalty0
  393--402, 2007.

\bibitem[Huang et~al.(2018)Huang, Slaney, and Elhilali]{huang2018connecting}
Nicholas Huang, Malcolm Slaney, and Mounya Elhilali.
\newblock Connecting deep neural networks to physical, perceptual, and
  electrophysiological auditory signals.
\newblock \emph{Frontiers in neuroscience}, 12:\penalty0 532, 2018.

\bibitem[Humphreys(1939)]{humphreys_acquisition_1939}
L.~G. Humphreys.
\newblock Acquisition and extinction of verbal expectations in a situation
  analogous to conditioning.
\newblock \emph{Journal of Experimental Psychology}, 25:\penalty0 294--301,
  1939.
\newblock ISSN 0022-1015.
\newblock \doi{10.1037/h0053555}.
\newblock Place: US Publisher: American Psychological Association.

\bibitem[Huth et~al.(2016)Huth, de~Heer, Griffiths, Theunissen, and
  Gallant]{huth_natural_2016}
Alexander~G. Huth, Wendy~A. de~Heer, Thomas~L. Griffiths, Frédéric~E.
  Theunissen, and Jack~L. Gallant.
\newblock Natural speech reveals the semantic maps that tile human cerebral
  cortex.
\newblock \emph{Nature}, 532\penalty0 (7600):\penalty0 453--458, April 2016.
\newblock ISSN 0028-0836, 1476-4687.
\newblock \doi{10.1038/nature17637}.

\bibitem[Jain and Huth(2018)]{jain2018incorporating}
Shailee Jain and Alexander Huth.
\newblock Incorporating context into language encoding models for fmri.
\newblock \emph{Advances in neural information processing systems}, 31, 2018.

\bibitem[Kell and McDermott(2019)]{kell2019deep}
Alexander~JE Kell and Josh~H McDermott.
\newblock Deep neural network models of sensory systems: windows onto the role
  of task constraints.
\newblock \emph{Current opinion in neurobiology}, 55:\penalty0 121--132, 2019.

\bibitem[Kell et~al.(2018)Kell, Yamins, Shook, Norman-Haignere, and
  McDermott]{kell2018task}
Alexander~JE Kell, Daniel~LK Yamins, Erica~N Shook, Sam~V Norman-Haignere, and
  Josh~H McDermott.
\newblock A task-optimized neural network replicates human auditory behavior,
  predicts brain responses, and reveals a cortical processing hierarchy.
\newblock \emph{Neuron}, 98\penalty0 (3):\penalty0 630--644, 2018.

\bibitem[Kellis et~al.(2010)Kellis, Miller, Thomson, Brown, House, and
  Greger]{kellis2010decoding}
Spencer Kellis, Kai Miller, Kyle Thomson, Richard Brown, Paul House, and
  Bradley Greger.
\newblock Decoding spoken words using local field potentials recorded from the
  cortical surface.
\newblock \emph{Journal of neural engineering}, 7\penalty0 (5):\penalty0
  056007, 2010.

\bibitem[Kietzmann et~al.(2018)Kietzmann, McClure, and
  Kriegeskorte]{kietzmann2018deep}
Tim~C Kietzmann, Patrick McClure, and Nikolaus Kriegeskorte.
\newblock Deep neural networks in computational neuroscience.
\newblock \emph{BioRxiv}, page 133504, 2018.

\bibitem[Koumura et~al.(2019)Koumura, Terashima, and
  Furukawa]{koumura2019cascaded}
Takuya Koumura, Hiroki Terashima, and Shigeto Furukawa.
\newblock Cascaded tuning to amplitude modulation for natural sound
  recognition.
\newblock \emph{Journal of Neuroscience}, 39\penalty0 (28):\penalty0
  5517--5533, 2019.

\bibitem[Kriegeskorte(2015)]{kriegeskorte2015deep}
Nikolaus Kriegeskorte.
\newblock Deep neural networks: a new framework for modeling biological vision
  and brain information processing.
\newblock \emph{Annual review of vision science}, 1:\penalty0 417--446, 2015.

\bibitem[Kuhl et~al.(2005)Kuhl, Conboy, Padden, Nelson, and
  Pruitt]{kuhl2005early}
Patricia~K Kuhl, Barbara~T Conboy, Denise Padden, Tobey Nelson, and Jessica
  Pruitt.
\newblock Early speech perception and later language development: Implications
  for the" critical period".
\newblock \emph{Language learning and development}, 1\penalty0 (3-4):\penalty0
  237--264, 2005.

\bibitem[Lakretz et~al.(2021)Lakretz, Desbordes, King, Crabbé, Oquab, and
  Dehaene]{lakretz_can_2021}
Yair Lakretz, Théo Desbordes, Jean-Rémi King, Benoît Crabbé, Maxime Oquab,
  and Stanislas Dehaene.
\newblock Can {RNNs} learn {Recursive} {Nested} {Subject}-{Verb} {Agreements}?
\newblock \emph{arXiv:2101.02258 [cs]}, January 2021.
\newblock URL \url{http://arxiv.org/abs/2101.02258}.
\newblock arXiv: 2101.02258.

\bibitem[Lerner et~al.(2011)Lerner, Honey, Silbert, and
  Hasson]{lerner_topographic_2011}
Y.~Lerner, C.~J. Honey, L.~J. Silbert, and U.~Hasson.
\newblock Topographic {Mapping} of a {Hierarchy} of {Temporal} {Receptive}
  {Windows} {Using} a {Narrated} {Story}.
\newblock \emph{Journal of Neuroscience}, 31\penalty0 (8):\penalty0 2906--2915,
  February 2011.
\newblock ISSN 0270-6474, 1529-2401.
\newblock \doi{10.1523/JNEUROSCI.3684-10.2011}.

\bibitem[Li et~al.(2021)Li, Bhattasali, Zhang, Franzluebbers, Luh, Spreng,
  Brennan, Yang, Pallier, and Hale]{li2021petit}
Jixing Li, Shohini Bhattasali, Shulin Zhang, Berta Franzluebbers, Wen-Ming Luh,
  R~Nathan Spreng, Jonathan~R Brennan, Yiming Yang, Christophe Pallier, and
  John~T Hale.
\newblock Le petit prince: A multilingual fmri corpus using ecological stimuli.
\newblock \emph{bioRxiv}, 2021.

\bibitem[Malik-Moraleda et~al.(2022)Malik-Moraleda, Ayyash, Gallée, Affourtit,
  Hoffmann, Mineroff, Jouravlev, and
  Fedorenko]{malik-moraleda_investigation_2022}
Saima Malik-Moraleda, Dima Ayyash, Jeanne Gallée, Josef Affourtit, Malte
  Hoffmann, Zachary Mineroff, Olessia Jouravlev, and Evelina Fedorenko.
\newblock An investigation across 45 languages and 12 language families reveals
  a universal language network.
\newblock \emph{Nature Neuroscience}, 25\penalty0 (8):\penalty0 1014--1019,
  August 2022.
\newblock ISSN 1546-1726.
\newblock \doi{10.1038/s41593-022-01114-5}.
\newblock Number: 8 Publisher: Nature Publishing Group.

\bibitem[Mesgarani et~al.(2014)Mesgarani, Cheung, Johnson, and
  Chang]{mesgarani_phonetic_2014}
Nima Mesgarani, Connie Cheung, Keith Johnson, and Edward~F. Chang.
\newblock Phonetic {Feature} {Encoding} in {Human} {Superior} {Temporal}
  {Gyrus}.
\newblock \emph{Science}, 343\penalty0 (6174):\penalty0 1006--1010, February
  2014.
\newblock ISSN 0036-8075, 1095-9203.
\newblock \doi{10.1126/science.1245994}.
\newblock Publisher: American Association for the Advancement of Science
  Section: Report.

\bibitem[Millet and Dunbar(2022)]{millet2022self}
Juliette Millet and Ewan Dunbar.
\newblock Do self-supervised speech models develop human-like perception
  biases?
\newblock \emph{AAAI 2022 Workshop, Self-Supervised Learning for Audio and
  Speech Processing}, 2022.

\bibitem[Millet and King(2021)]{millet2021inductive}
Juliette Millet and Jean-Remi King.
\newblock Inductive biases, pretraining and fine-tuning jointly account for
  brain responses to speech.
\newblock \emph{arXiv preprint arXiv:2103.01032}, 2021.

\bibitem[Millet et~al.(2021)Millet, Chitoran, and
  Dunbar]{millet-etal-2021-predicting}
Juliette Millet, Ioana Chitoran, and Ewan Dunbar.
\newblock Predicting non-native speech perception using the perceptual
  assimilation model and state-of-the-art acoustic models.
\newblock In \emph{Proceedings of the 25th Conference on Computational Natural
  Language Learning}, pages 661--673, Online, November 2021. Association for
  Computational Linguistics.
\newblock \doi{10.18653/v1/2021.conll-1.51}.

\bibitem[Mitchell et~al.(2008)Mitchell, Shinkareva, Carlson, Chang, Malave,
  Mason, and Just]{mitchell2008predicting}
Tom~M Mitchell, Svetlana~V Shinkareva, Andrew Carlson, Kai-Min Chang, Vicente~L
  Malave, Robert~A Mason, and Marcel~Adam Just.
\newblock Predicting human brain activity associated with the meanings of
  nouns.
\newblock \emph{science}, 320\penalty0 (5880):\penalty0 1191--1195, 2008.

\bibitem[Mugler et~al.(2014)Mugler, Patton, Flint, Wright, Schuele, Rosenow,
  Shih, Krusienski, and Slutzky]{mugler2014direct}
Emily~M Mugler, James~L Patton, Robert~D Flint, Zachary~A Wright, Stephan~U
  Schuele, Joshua Rosenow, Jerry~J Shih, Dean~J Krusienski, and Marc~W Slutzky.
\newblock Direct classification of all american english phonemes using signals
  from functional speech motor cortex.
\newblock \emph{Journal of neural engineering}, 11\penalty0 (3):\penalty0
  035015, 2014.

\bibitem[Naselaris et~al.(2011)Naselaris, Kay, Nishimoto, and
  Gallant]{naselaris2011encoding}
Thomas Naselaris, Kendrick~N Kay, Shinji Nishimoto, and Jack~L Gallant.
\newblock Encoding and decoding in fmri.
\newblock \emph{Neuroimage}, 56\penalty0 (2):\penalty0 400--410, 2011.

\bibitem[Nastase et~al.(2020)Nastase, Liu, Hillman, Zadbood, Hasenfratz,
  Keshavarzian, Chen, Honey, Yeshurun, Regev, et~al.]{nastase2020narratives}
Samuel~A Nastase, Yun-Fei Liu, Hanna Hillman, Asieh Zadbood, Liat Hasenfratz,
  Neggin Keshavarzian, Janice Chen, Christopher~J Honey, Yaara Yeshurun, Mor
  Regev, et~al.
\newblock Narratives: fmri data for evaluating models of naturalistic language
  comprehension. preprint.
\newblock \emph{Neuroscience, December}, pages 2020--06, 2020.

\bibitem[Panayotov et~al.(2015)Panayotov, Chen, Povey, and
  Khudanpur]{panayotov2015librispeech}
Vassil Panayotov, Guoguo Chen, Daniel Povey, and Sanjeev Khudanpur.
\newblock Librispeech: an asr corpus based on public domain audio books.
\newblock In \emph{2015 IEEE international conference on acoustics, speech and
  signal processing (ICASSP)}, pages 5206--5210. IEEE, 2015.

\bibitem[Pasad et~al.(2021)Pasad, Chou, and Livescu]{pasad2021layer}
Ankita Pasad, Ju-Chieh Chou, and Karen Livescu.
\newblock Layer-wise analysis of a self-supervised speech representation model.
\newblock \emph{arXiv preprint arXiv:2107.04734}, 2021.

\bibitem[Pedregosa et~al.(2011)Pedregosa, Varoquaux, Gramfort, Michel, Thirion,
  Grisel, Blondel, Prettenhofer, Weiss, Dubourg, et~al.]{pedregosa2011scikit}
Fabian Pedregosa, Ga{\"e}l Varoquaux, Alexandre Gramfort, Vincent Michel,
  Bertrand Thirion, Olivier Grisel, Mathieu Blondel, Peter Prettenhofer, Ron
  Weiss, Vincent Dubourg, et~al.
\newblock Scikit-learn: Machine learning in python.
\newblock \emph{the Journal of machine Learning research}, 12:\penalty0
  2825--2830, 2011.

\bibitem[Pereira et~al.(2018)Pereira, Lou, Pritchett, Ritter, Gershman,
  Kanwisher, Botvinick, and Fedorenko]{pereira2018toward}
Francisco Pereira, Bin Lou, Brianna Pritchett, Samuel Ritter, Samuel~J
  Gershman, Nancy Kanwisher, Matthew Botvinick, and Evelina Fedorenko.
\newblock Toward a universal decoder of linguistic meaning from brain
  activation.
\newblock \emph{Nature communications}, 9\penalty0 (1):\penalty0 1--13, 2018.

\bibitem[Petkov et~al.(2015)Petkov, Kikuchi, Milne, Mishkin, Rauschecker, and
  Logothetis]{petkov_different_2015}
Christopher~I. Petkov, Yukiko Kikuchi, Alice~E. Milne, Mortimer Mishkin,
  Josef~P. Rauschecker, and Nikos~K. Logothetis.
\newblock Different forms of effective connectivity in primate frontotemporal
  pathways.
\newblock \emph{Nature Communications}, 6\penalty0 (1):\penalty0 6000, January
  2015.
\newblock ISSN 2041-1723.
\newblock \doi{10.1038/ncomms7000}.
\newblock Number: 1 Publisher: Nature Publishing Group.

\bibitem[Poeppel et~al.(2012)Poeppel, Emmorey, Hickok, and
  Pylkk{\"a}nen]{poeppel2012towards}
David Poeppel, Karen Emmorey, Gregory Hickok, and Liina Pylkk{\"a}nen.
\newblock Towards a new neurobiology of language.
\newblock \emph{Journal of Neuroscience}, 32\penalty0 (41):\penalty0
  14125--14131, 2012.

\bibitem[Qian et~al.(2016)Qian, Qiu, and Huang]{qian2016bridging}
Peng Qian, Xipeng Qiu, and Xuanjing Huang.
\newblock Bridging lstm architecture and the neural dynamics during reading.
\newblock \emph{arXiv preprint arXiv:1604.06635}, 2016.

\bibitem[Radford et~al.(2019)Radford, Wu, Child, Luan, Amodei, Sutskever,
  et~al.]{gpt2}
Alec Radford, Jeffrey Wu, Rewon Child, David Luan, Dario Amodei, Ilya
  Sutskever, et~al.
\newblock Language models are unsupervised multitask learners.
\newblock \emph{OpenAI blog}, 1\penalty0 (8):\penalty0 9, 2019.

\bibitem[Schatz(2016)]{schatz2016abx}
Thomas Schatz.
\newblock \emph{ABX-discriminability measures and applications}.
\newblock PhD thesis, Universit{\'e} Paris 6 (UPMC), 2016.

\bibitem[Schrimpf et~al.(2021)Schrimpf, Blank, Tuckute, Kauf, Hosseini,
  Kanwisher, Tenenbaum, and Fedorenko]{schrimpf2021neural}
Martin Schrimpf, Idan~Asher Blank, Greta Tuckute, Carina Kauf, Eghbal~A
  Hosseini, Nancy Kanwisher, Joshua~B Tenenbaum, and Evelina Fedorenko.
\newblock The neural architecture of language: Integrative modeling converges
  on predictive processing.
\newblock \emph{Proceedings of the National Academy of Sciences}, 118\penalty0
  (45), 2021.

\bibitem[Schwartz et~al.(2019)Schwartz, Toneva, and
  Wehbe]{schwartz2019inducing}
Dan Schwartz, Mariya Toneva, and Leila Wehbe.
\newblock Inducing brain-relevant bias in natural language processing models.
\newblock \emph{Advances in neural information processing systems}, 32, 2019.

\bibitem[Shamma et~al.(2021)Shamma, Patel, Mukherjee, Marion, Khalighinejad,
  Han, Herrero, Bickel, Mehta, and Mesgarani]{shamma2021learning}
Shihab Shamma, Prachi Patel, Shoutik Mukherjee, Guilhem Marion, Bahar
  Khalighinejad, Cong Han, Jose Herrero, Stephan Bickel, Ashesh Mehta, and Nima
  Mesgarani.
\newblock Learning speech production and perception through sensorimotor
  interactions.
\newblock \emph{Cerebral cortex communications}, 2\penalty0 (1):\penalty0
  tgaa091, 2021.

\bibitem[Stephenson et~al.(2019)Stephenson, Feather, Padhy, Elibol, Tang,
  McDermott, and Chung]{stephenson2019untangling}
Cory Stephenson, Jenelle Feather, Suchismita Padhy, Oguz Elibol, Hanlin Tang,
  Josh McDermott, and SueYeon Chung.
\newblock Untangling in invariant speech recognition.
\newblock \emph{Advances in neural information processing systems}, 32, 2019.

\bibitem[Thompson et~al.(2021)Thompson, Bengio, Formisano, and
  Sch{\"o}nwiesner]{thompson2021training}
Jessica~AF Thompson, Yoshua Bengio, Elia Formisano, and Marc Sch{\"o}nwiesner.
\newblock Training neural networks to recognize speech increased their
  correspondence to the human auditory pathway but did not yield a shared
  hierarchy of acoustic features.
\newblock \emph{bioRxiv}, 2021.

\bibitem[Toneva and Wehbe(2019)]{toneva_interpreting_2019}
Mariya Toneva and Leila Wehbe.
\newblock Interpreting and improving natural-language processing (in machines)
  with natural language-processing (in the brain).
\newblock \emph{arXiv:1905.11833 [cs, q-bio]}, November 2019.
\newblock arXiv: 1905.11833.

\bibitem[Vaidya et~al.(2022)Vaidya, Jain, and Huth]{vaidya2022self}
Aditya~R Vaidya, Shailee Jain, and Alexander~G Huth.
\newblock Self-supervised models of audio effectively explain human cortical
  responses to speech.
\newblock \emph{arXiv preprint arXiv:2205.14252}, 2022.

\bibitem[Wang et~al.(2021)Wang, Riviere, Lee, Wu, Talnikar, Haziza, Williamson,
  Pino, and Dupoux]{wang2021voxpopuli}
Changhan Wang, Morgane Riviere, Ann Lee, Anne Wu, Chaitanya Talnikar, Daniel
  Haziza, Mary Williamson, Juan Pino, and Emmanuel Dupoux.
\newblock Voxpopuli: A large-scale multilingual speech corpus for
  representation learning, semi-supervised learning and interpretation.
\newblock \emph{arXiv preprint arXiv:2101.00390}, 2021.

\bibitem[Weerts et~al.(2021)Weerts, Rosen, Clopath, and
  Goodman]{weerts2021psychometrics}
Lotte Weerts, Stuart Rosen, Claudia Clopath, and Dan~FM Goodman.
\newblock The psychometrics of automatic speech recognition.
\newblock \emph{bioRxiv}, 2021.

\bibitem[Xu et~al.(2020)Xu, Baevski, Likhomanenko, Tomasello, Conneau,
  Collobert, Synnaeve, and Auli]{xu2021self}
Qiantong Xu, Alexei Baevski, Tatiana Likhomanenko, Paden Tomasello, Alexis
  Conneau, Ronan Collobert, Gabriel Synnaeve, and Michael Auli.
\newblock Self-training and {Pre}-training are {Complementary} for {Speech}
  {Recognition}, October 2020.
\newblock arXiv:2010.11430 [cs, eess].

\bibitem[Yamins and DiCarlo(2016)]{yamins2016using}
Daniel~LK Yamins and James~J DiCarlo.
\newblock Using goal-driven deep learning models to understand sensory cortex.
\newblock \emph{Nature neuroscience}, 19\penalty0 (3):\penalty0 356--365, 2016.

\end{thebibliography}
\bibliographystyle{plainnat}




\clearpage
\appendix
\setcounter{figure}{0}
\renewcommand{\thefigure}{S\arabic{figure}}
\renewcommand{\thetable}{S\arabic{table}}

\section{Appendix}
\subsection{Self-supervised loss formula}
\label{ap:loss_self}
Wav2vec 2.0, when trained in a self-supervised way, uses a loss ($L$) which is the weighted combination of two losses: one diversity loss ($L_d$), which pushes the quantization module to contain representations that are as diverse as possible, and one Contrastive Predictive Coding loss ($L_m$), which pushes the model to choose, from the context network output $c$, the right quantized representation ($q$) of some masked input, among other possible representations. $L_m$ has the following formula, for some masked time step $t$:
\begin{align}
    \mathcal{L}_{m}=-\log \frac{\exp \left(\operatorname{sim}\left(\mathbf{c}_{t}, \mathbf{q}_{t}\right) / \kappa\right)}{\sum_{\tilde{\mathbf{q}} \sim \mathbf{Q}_{t}} \exp \left(\operatorname{sim}\left(\mathbf{c}_{t}, \tilde{\mathbf{q}}\right) / \kappa\right)}
\end{align}
with $\operatorname{sim}(\mathbf{a}, \mathbf{b})=\mathbf{a}^{T} \mathbf{b} /\|\mathbf{a}\|\|\mathbf{b}\|$, $\kappa$ the temperature, which is constant during training, $Q_t$ the set of $K+1$ quantized candidate the model has to choose from, including the right one, i.e. $q_t$.

$L_d$ is included to encourage the equal use of the $V$ possible entries of each of the $G$ codebooks of the quantization module. The goal is to maximize the entropy of the averaged softmax distribution over the codebook entries for each codebook $\bar{p}_{g}$, across a set of utterances:
\begin{align}
    \mathcal{L}_{d}=\frac{1}{G V} \sum_{g=1}^{G}-H\left(\bar{p}_{g}\right)=\frac{1}{G V} \sum_{g=1}^{G} \sum_{v=1}^{V} \bar{p}_{g, v} \log \bar{p}_{g, v}
\end{align}

\subsection{Supervised loss formula}
\label{ap:loss_sup}
When trained in a supervised way, wav2vec 2.0 is trained to optimise a Connectionist Temporal Classification loss parameterized over $\theta$:
\begin{align}
    \mathrm{argmin}_{\theta} \ -\log \sum_{a \in a_{U,V}} \prod_{t=1}^{d_t} p_{\mathrm{CTC}}\left(a_{t} \mid m_\theta(U)\right) \enspace ,
\end{align}
where $m_\theta(U) \in \mathbb{R}^{d_\tau \times d_v}$ are the probabilistic predictions of the model at each $\tau$ time sample given the input raw waveform $U\in \mathbb{R}^{d_\tau \times d_u}$, $V \in \mathbb{R}^{d_t \times d_v}$ are the true transcriptions of $U$, and $a_{U,V}$ is the set of all possible alignments between $U$ and $V$.

\subsection{Preprocessing of the model's activations}
\label{ap:preproc}
The activations of the network $X \in \mathbb{R}^{d_{\hat{t}} \times d_x}$ are first normalized to be between $[0, 1]$ for each listening session. Then, we use nistats \citep{nilearn} \texttt{compute\_regressor} function with the `glover' model to temporally convolve ($h \in \mathbb{R}^{d_{\hat{t}}}$) and temporally down-sample ( using $g: \mathbb{R}^{d_{\hat{t}}} \rightarrow \mathbb{R}^{d_t}$) each artificial neuron $j$:
\begin{align}
    \hat{x}^{(j)} = g \Big ( x^{(j)} * h \Big ) \enspace .
\end{align}

\subsection{Penalized linear model - Ridge regression}
\label{ap:ridge}
For each split $s$, we fit an $\ell_2$-penalized linear model $V \in \mathbb{R}^{d_x\times d_z}$ trained to predict the transformed BOLD time series from the model activations for each dimension independently. The formula of the optimization is the following:
\begin{align}
    \mathrm{argmin}_{V} \sum_{i \in \mathrm{train}_s} (V^\top \hat{X}_{i} - y_{i})^2 + \lambda \|V\|^2
    \enspace .
\end{align}

\subsection{Probing the linguistic features encoded in wav2vec2 activations}\label{ap:probs}

\begin{figure}[ht]
  \centering
  \includegraphics[width=\textwidth]{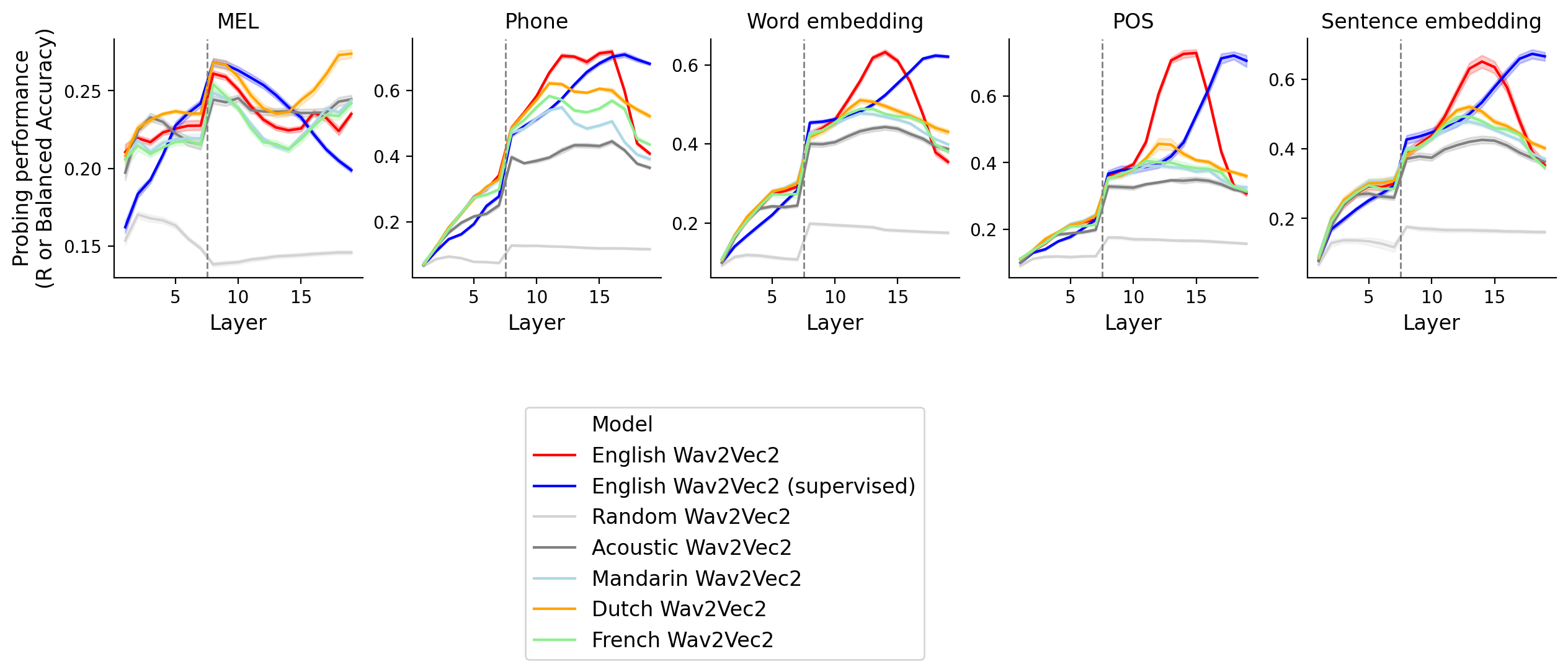}
  \caption{\textbf{Linguistic features encoded in each layer of the networks.}
  For each layer of each network, we train a l2-penalized linear model from scikit-learn \citep{pedregosa2011scikit} to predict several linguistic categories given the embedding. The tested categories are the following: MEL (the MEL spectrogram of the audio, d=128), phone (the phoneme, categorical, d=39), the word embedding of the word (computed with spaCy (\protect\url{https://spacy.io}) English model, d=300), the Part-Of-Speech (POS) of the word provided by spaCy (categorical feature, n=19), and the embedding of the sentence, computed using Laser (\protect\url{https://github.com/facebookresearch/LASER}) (d=1,024). We train and test the linear probe on a subset of Timit data (\protect\url{https://catalog.ldc.upenn.edu/LDC93s1}), using a 10-folds cross-validation scheme, and report the probing accuracy (either $R$ for continuous variables or balanced accuracy for categorical variables) for each possible target feature. We average the corresponding probing performances across the 10 folds. Error bars are standard errors of the mean across folds.}
\label{fig-probe}
\end{figure}

\revised{Interpreting the representations of deep learning models is notoriously difficult. To address this issue, \citep{pasad2021layer} explored the encoding of local acoustic features, phone identity, word identity and word meaning across layers. Similarly, \citep{millet-etal-2021-predicting} compared representations to human behavioural data to assess whether they better captured listeners' perception of higher-level phonemic properties or of lower-level subphonemic properties of speech stimuli. Finally, \citep{vaidya2022self} recent study explores filter banks, spectrograms, phonemes and words across layers. Here, we complement these analyses by showing that self-supervised learning allows wav2vec 2.0 to learn represents, along its hierarchy the representations of MEL spectrograms, phonetic categories and word embeddings (Figure \ref{fig-probe}).}

For this, we perform a ridge regression on the Timit dataset\footnote{https://catalog.ldc.upenn.edu/LDC93S1} to predict five auditory and linguistic features from the activation functions of each layer and model of the present paper. We study the following features: 
\begin{itemize}
    \item the MEL spectrogram of the audio, computed using librosa (d=128)
    \item the phonemes (categorical features). We use the transcripts and alignments provided in Timit. 
    \item the word embedding and part-of-speech of the words. The time alignments for words are provided by Timit. We use spaCy to compute the word embedding (medium model, d=300), and their part-of-speech (categorical feature, d=19). 
    \item the sentence embedding of each sample, provided by Laser. 
\end{itemize}
We use a subset of 1,680 samples from Timit, each sample being an audio recording of a short sentence (<10 seconds) from 24 speakers. The model's activations were mean-pooled to the sampling rate of each feature. 

The results show that the layers of wav2vec 2.0 partially follow the hierarchy predicted from neuro-linguistics \citep{hickok2007cortical} (Table \ref{tab:probe}): the first layers of the transformer best account for the spectro-temporal information, whereas deeper layers best account for the phonetic, word-level and sentence level information. While all of these features emerge with training (Figure \ref{fig-probe}), only the highest-level features (phone, word and sentence-level) appear to be specific to speech and to the language with which wav2vec 2.0 was trained (Figure \ref{fig-probe}).

\revised{
Interestingly, the word and sentence-level features are encoded deeper in the supervised network (best layer=18 in Table \ref{tab:probe}) compared to the unsupervised network (best layer=14), which suggests that self-supervised learning generates a reservoir representations in its middle layers, reservoir which may partly overlap with the labels used in supervised learning. Together with our ABX tests, and layer-wise tuning of each voxel (Figure \ref{fig-hierarchy}), these elements suggest that the representations of speech shaped by our experience are learnt and instantiated in the superior temporal gyrus and sulcus. These elements, \revised{consistent with previous electrophysiological studies \citep{mesgarani_phonetic_2014}}, thus provide a coherent spectrum of evidence for the location of acquired speech representations in the brain.}

\subsection{Noise ceiling analysis}\label{ap:noise_ceil}

The noise in fMRI recordings is inevitable. To estimate the maximum explainable signal given this level of noise, we follow previous studies and employ a shared-response model, or "noise ceiling" \citep{huth_natural_2016,caucheteux2022brains, caucheteux_gpt-2s_2021}. Precisely, we predict the brain signals of one subject given the brain activity of the other subjects, in response to the same audio recording. In practice, we apply the same evaluation as Equation \eqref{np_score}, for one subject $s$ and one voxel $v$, but we use the average brain signals of other subjects' brains $\overline{Y}^{(s)} = \frac{1}{|\mathcal{S}|} \sum_{s' \neq s} Y^{(s')}$ instead of the activations $X$. As a result, the "noise ceiling" of one subject ($s$) and one voxel ($v$) is computed as follows:
\begin{equation}\label{equ_noiseceil}
 R_{\mathrm{noiseceil}} = \mathrm{Corr} 
\big( 
 W \cdot \overline{Y}^{(s)}, 
  Y^{(s,v)}
  \big) \quad , 
\end{equation}
where $W$ is an $\ell_2$-penalized linear regression fitted on separate train data, using a cross validation setting with five test folds. 

\begin{figure}[ht]
  \centering
  \includegraphics[width=0.94\textwidth]{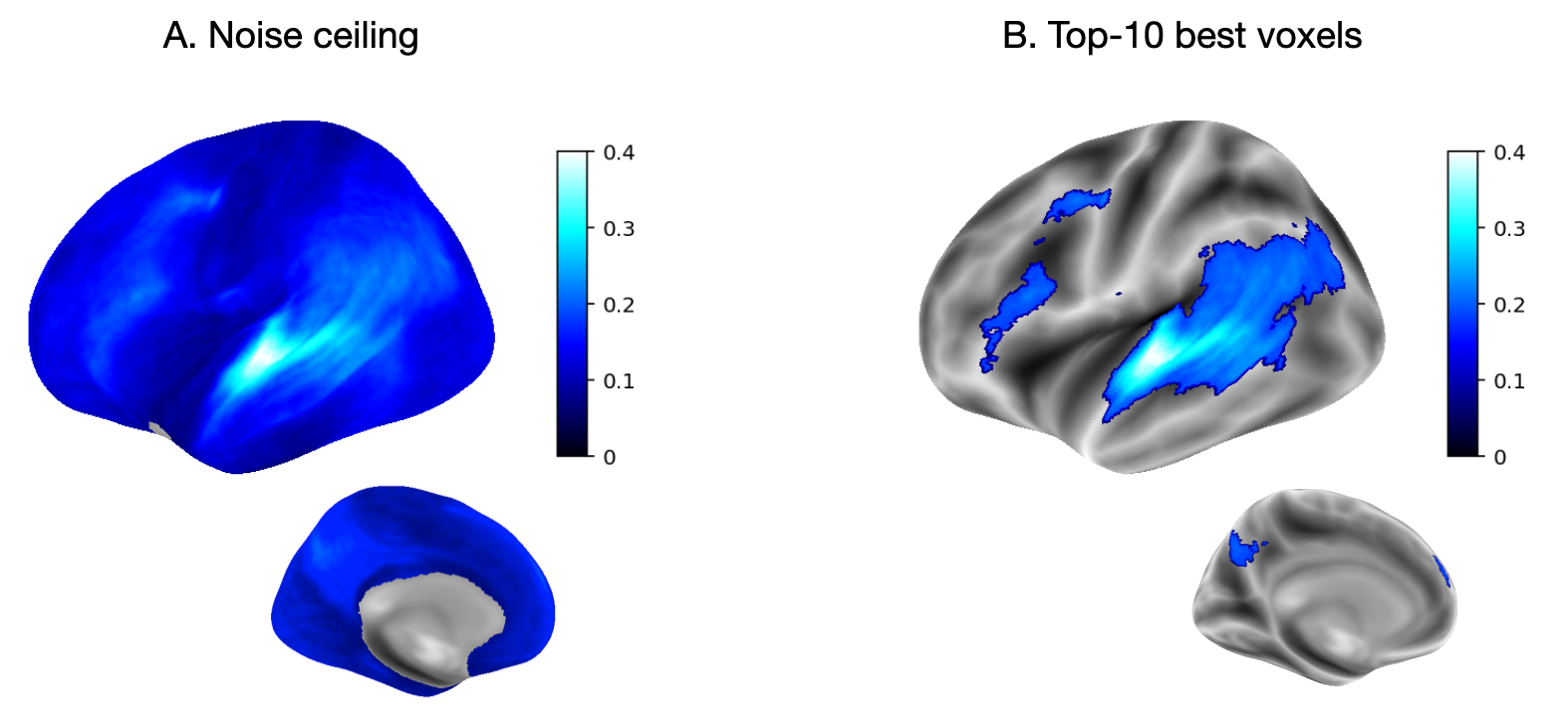}
  \caption{\textbf{Noise ceiling.}  \textbf{A.} Noise ceiling estimates computed on 290 subjects of the Narratives dataset, averaged across subject. We only display the significant voxels across subjects ($p<10^{-18}$). \textbf{B.} Same as A, but we only display the 10\% voxels with the best noise ceiling estimates on average across subjects. 
}
\label{fig-noiseceil}
\end{figure}

We compute such noise ceiling on 290 subjects of the Narrative dataset listening to the same stories (Figure \ref{fig-noiseceil}). We report the noise ceiling across voxels in Figure \ref{fig-noiseceil}, and, in Table \ref{tab:np_noiseceil}, the \revised{brain scores} of the networks studied in the main paper normalised by the noise ceiling. Precisely, for each voxel, we divide the average \revised{brain scores} by the noise ceiling for this particular voxel. While low on average, the unsupervised wav2vec2 model reaches 74\% of the noise ceiling in Heschel, and more than 20\% in STS, STS and IFG. 

\begin{table}[ht]
    \centering

\begin{tabular}{lrrrrrrr}
\toprule
{} &     Average &   Top10 &  Heschl &     STG &     STS &     IFG &   Motor \\
\midrule
Random wav2vec2      &   13.9\% &   29.0\% &   66.9\% &   32.0\% &   21.8\% &   15.9\% &   11.9\% \\
Non-Speech           &   16.4\% &   33.9\% &   71.0\% &   36.8\% &   26.9\% &   19.0\% &   11.7\% \\
Non-Native           &   17.6\% &   35.9\% &   73.0\% &   39.0\% &   29.1\% &   21.0\% &   12.9\% \\
Native, Supervised   &   18.3\% &   36.7\% &   74.2\% &   39.6\% &   29.8\% &   21.2\% &   13.6\% \\
Native, Unsupervised &   18.8\% &   37.9\% &   74.4\% &   40.3\% &   31.3\% &   22.8\% &   13.8\% \\
Noise ceiling        &  100.0\% &  100.0\% &  100.0\% &  100.0\% &  100.0\% &  100.0\% &  100.0\% \\
\bottomrule
\end{tabular} 
\caption{\textbf{\revised{Brain scores} \emph{with} noise ceiling normalisation}. \revised{Brain scores} divided by the noise ceiling, for the Narrative dataset, on average across all voxels (`Average'), for the 10\% best voxels of the noise ceiling (`Top10', Figure \ref{ap:noise_ceil}) and the voxels of five regions of interests.
}
    \label{tab:np_noiseceil} 
\end{table}

\begin{table}[ht]
    \centering
\begin{tabular}{lrrrrrrr}
\toprule
{} &    Average &  Top10 & Heschl &    STG &    STS &    IFG &  Motor \\
\midrule
Random wav2vec2      &  0.019 &  0.069 &  0.192 &  0.071 &  0.044 &  0.024 &  0.011 \\
Non-Speech           &  0.022 &  0.080 &  0.205 &  0.081 &  0.055 &  0.028 &  0.011 \\
Non-Native           &  0.024 &  0.085 &  0.211 &  0.086 &  0.059 &  0.031 &  0.012 \\
Native, Supervised   &  0.025 &  0.086 &  0.213 &  0.087 &  0.060 &  0.032 &  0.013 \\
Native, Unsupervised &  0.025 &  0.089 &  0.214 &  0.089 &  0.063 &  0.034 &  0.013 \\
Noise ceiling        &  0.117 &  0.219 &  0.287 &  0.181 &  0.196 &  0.149 &  0.094 \\
\bottomrule
\end{tabular}
\caption{\textbf{\revised{Brain scores} \emph{without} noise ceiling normalisation} Same as Table \ref{tab:np_noiseceil}, but without dividing by the noise ceiling estimates. 
}
\label{tab:np}
\end{table}

\begin{table}[ht]
    \centering
   \resizebox{0.98\textwidth}{!}{\begin{tabular}{llllllll}
\toprule
{} &             Avg &  Top10NoiseCeil &          Heschl &             STG &             STS &             IFG &           Motor \\
\midrule
Unsupervised  &  0.03 +/- 0.001 &  0.09 +/- 0.002 &  0.21 +/- 0.007 &  0.09 +/- 0.003 &  0.06 +/- 0.002 &  0.03 +/- 0.001 &  0.01 +/- 0.001 \\
Supervised    &  0.02 +/- 0.001 &  0.09 +/- 0.002 &  0.21 +/- 0.007 &  0.09 +/- 0.003 &  0.06 +/- 0.002 &  0.03 +/- 0.001 &  0.01 +/- 0.001 \\
Noise ceiling &  0.12 +/- 0.006 &  0.22 +/- 0.006 &  0.29 +/- 0.008 &  0.18 +/- 0.006 &  0.20 +/- 0.006 &  0.15 +/- 0.006 &  0.09 +/- 0.006 \\
Ratio         &  0.19 +/- 0.006 &  0.38 +/- 0.010 &  0.74 +/- 0.025 &  0.40 +/- 0.013 &  0.31 +/- 0.011 &  0.23 +/- 0.010 &  0.14 +/- 0.014 \\
\bottomrule
\end{tabular}}
    \caption{\revised{Brain scores} and noise ceiling estimates. Ratio indicate the unsupervised model divided by the noise ceiling. Scores are averaged across subjects and either all the voxels (`Avg') or the voxels of the selected regions of interests.}
    \label{tab:np_noiseceil_models}
\end{table}

Below, we report the \revised{brain scores} of our models, normalised by such noise ceiling. Precisely, we compute the \revised{brain scores} for each subject and voxels

\begin{figure}[ht]
  \centering
  \includegraphics[width=0.6\textwidth]{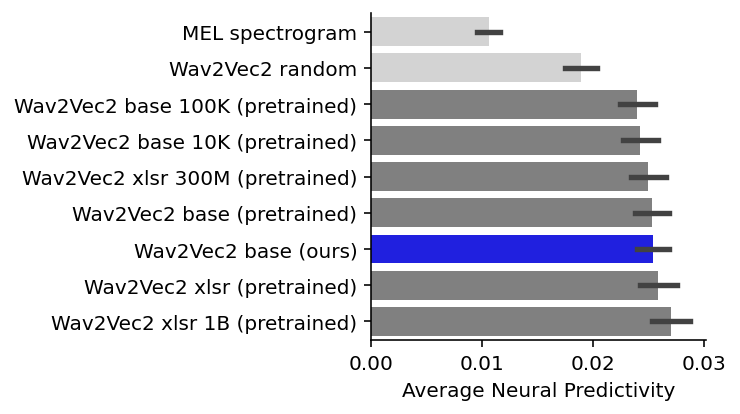}
  \caption{\textbf{\revised{Brain scores} of self-supervised pre-trained models.} \revised{Brain scores}, averaged across all voxels and subjects, for the MEL spectrogram, a wav2vec2 (base) architecture with random weights, wav2vec 2.0 (base) pre-trained with self supervised learning on 100K hours from Voxpopuli (Wang, 2021) (`wav2vec2-base-100k-voxpopuli’ from huggingface), on 10K hours from Voxpopuli (`wav2vec2-base-10k-voxpopuli’), on 53K hours of english (`wav2vec2-base`), two models pre-trained on the same multilingual corpus of 436K hours, with 300M (`wav2vec2-xls-r-300m`) and 1B parameters (`wav2vec2-xls-r-1b`), respectively, and our model trained on 600 hours of english speech (in blue). +/- refers to standard errors of the mean across subjects.  
}
\label{fig-pretrained}
\end{figure}

\begin{figure}[ht]
  \centering
  \includegraphics[width=\textwidth]{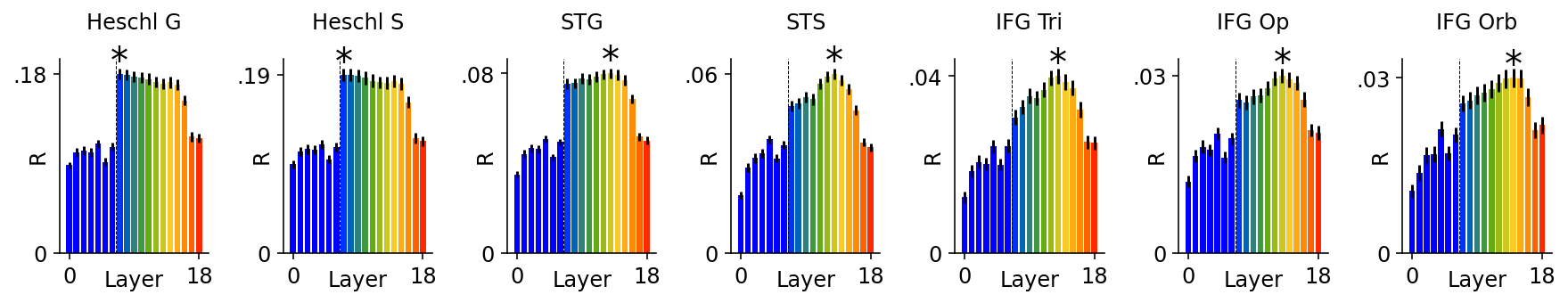}
  \caption{\textbf{\revised{Brain scores} for each layer of wav2vec 2.0} Same as figure 3B, but for different regions of the brain. \revised{Brain scores} are averaged across all voxels in each regions.  
}
\label{fig-layers}
\end{figure}

\begin{table}[ht]
    \centering
\resizebox{0.98\textwidth}{!}{\begin{tabular}{lrrrrrr}
\toprule
{} &   MEL &  Phone &  Wordemb &   POS &  Sentemb &   Average \\
\midrule
Random wav2vec2               &   2.0 &    8.7 &             8.0 &   8.9 &                 8.1 &   7.1 \\
Acoustic wav2vec2             &  12.5 &   15.7 &            14.0 &  14.4 &                14.2 &  14.2 \\
Mandarin wav2vec2             &   9.1 &   11.9 &            12.2 &  11.9 &                13.0 &  11.6 \\
French wav2vec2               &   8.0 &   11.0 &            12.7 &  11.8 &                13.0 &  11.3 \\
Dutch wav2vec2                &  18.9 &   11.4 &            12.0 &  12.4 &                13.0 &  13.5 \\
English wav2vec2              &   8.0 &   15.2 &            14.0 &  14.4 &                14.0 &  13.1 \\
English wav2vec2 (supervised) &   8.0 &   16.9 &            18.0 &  18.0 &                18.0 &  15.8 \\
Avg                           &   9.5 &   13.0 &            13.0 &  13.1 &                13.3 &  12.4 \\
\bottomrule
\end{tabular}}
    \caption{For each model (row) and target (column), the layer that maximizes probing performance (Figure S1), averaged across the 10 cross-validation folds.}
    \label{tab:probe}
\end{table}

\end{document}